\documentclass[prd,aps,showpacs,nofootinbib,superscriptaddress,tightenlines,amsmath,amssymb,eqsecnum]{revtex4}

\usepackage{graphicx}
\usepackage{bm}
\usepackage{epsfig}
\usepackage[latin1]{inputenc}
\usepackage{float,amsmath}

\def\lsim{\mathrel{\rlap{\lower4pt\hbox{$\sim$}}
    \raise1pt\hbox{$<$}}}                
\def\gsim{\mathrel{\rlap{\lower4pt\hbox{$\sim$}}
    \raise1pt\hbox{$>$}}}                
    
\newcommand{\beq}{\begin{equation}}
\newcommand{\eeq}{\end{equation}}
\newcommand{\bqa}{\begin{eqnarray}}
\newcommand{\eqa}{\end{eqnarray}}

\begin{document}
\title{Pre-equilibrium dilepton production from an anisotropic quark-gluon plasma}

\author{Mauricio Martinez}
\affiliation{Helmholtz Research School and Otto Stern School,\\
  Goethe-Universit\"at Frankfurt am Main, Ruth-Moufang-Str. 1,
  60438 Frankfurt am Main, Germany \\}
\author{Michael Strickland}
\affiliation{Institute f\"ur Theoretische Physik and Frankfurt Institute for Advanced Studies, \\
  Goethe-Universit\"at Frankfurt am Main, Max-von-Laue-Stra\ss{}e 1,
  D-60438 Frankfurt am Main, Germany \\}
\affiliation{Physics Department, Gettysburg College, Gettysburg, PA 17325, USA}

\begin {abstract}%
	{%
We calculate leading-order dilepton yields from a quark-gluon plasma 
which has a time-dependent anisotropy in momentum space.  Such 
anisotropies can arise during the earliest stages of quark-gluon 
plasma evolution due to the rapid longitudinal expansion of the 
created matter. Two phenomenological models for the proper time 
dependence of the parton hard momentum scale, $p_{\rm hard}$, and the 
plasma anisotropy parameter, $\xi$, are constructed which describe the 
transition of the plasma from its initial non-equilibrium state to an 
isotropic thermalized state. The first model constructed interpolates 
between 1+1 dimensional free streaming at early times and 1+1 
dimensional ideal hydrodynamical expansion at late times.  In the 
second model we include the effect of collisional broadening of the 
parton distribution functions in the early-time pre-equilibrium stage 
of plasma evolution. We find for both cases that for fixed initial 
conditions high-energy dilepton production is enhanced by 
pre-equilibrium emission. When the models are constrained to fixed 
final pion multiplicity the dependence of the resulting spectra on the 
assumed plasma isotropization time is reduced.  Using our most 
realistic collisionally-broadened model we find that high-transverse 
momentum dilepton production would be enhanced by at most 40\% at the 
Relativistic Heavy Ion Collider and 50\% at CERN Large Hadron Collider
if one assumes an isotropization/thermalization time 
of 2 fm/c. Given sufficiently precise experimental data this 
enhancement could be used to determine the plasma isotropization time 
experimentally.
	}
\end {abstract}
\pacs{11.15Bt, 04.25.Nx, 11.10Wx, 12.38Mh}\maketitle 


\section {Introduction}
\label{sec:intro}

With the ongoing ultrarelativistic heavy ion collision experiments at 
RHIC and future experiments planned at LHC the goal is to produce a 
deconfined plasma of quarks and gluons (QGP). This new state of matter 
is expected to be formed once the temperature of nuclear matter 
exceeds the critical temperature, $T_C\sim$ 200 MeV. However, many 
properties of the expected QGP are still poorly understood. One of the 
most difficult problems is the determination of the isotropization and 
thermalization time of the QGP, $\tau_{\rm iso}$ and $\tau_{\rm 
therm}$, respectively.\footnote{For simplicity from here forward we 
will assume that these two time scales are the same, $\tau_{\rm therm} 
= \tau_{\rm iso}$, so that the system achieves isotropization and 
thermalization at the same proper time given by $\tau_{\rm iso}$.} At 
RHIC energies it was found that, for $p_T\lesssim$ 2 GeV, the plasma 
elliptic flow, $v_2$, was well described by ideal hydrodynamical 
models. Based on early studies 
\cite{Huovinen:2001cy,Hirano:2002ds,Tannenbaum:2006ch} ideal 
hydrodynamical fits to elliptic flow data indicated that the matter 
can be modeled as a nearly-perfect fluid starting at extremely early 
times after the collision, $\tau_{\rm iso} \sim$ 0.6 fm/c 
\cite{Huovinen:2001cy}. However, recent hydrodynamical studies 
\cite{Luzum:2008cw} which include the effect of all 2nd-order 
transport coefficients consistent with conformal symmetry have shown 
that these initial estimates for the isotropization/thermalization 
time of the plasma have a sizable uncertainty due to poor knowledge of 
the proper initial conditions (CGC versus Glauber), details of plasma 
hadronization and subsequent hadronic cascade, etc.  As a result, it 
now seems that isotropization times up to $\tau_{\rm iso} \sim$ 2 fm/c 
are not ruled out by RHIC data and in order to further constrain this 
time additional theoretical and experimental input will be required.

As mentioned above there are uncertainties introduced in hydrodynamic 
modeling of the quark-gluon plasma (QGP) due primarily to the 
dependence of the results on the assumed initial conditions: initial 
energy density, spatial profile, flow velocity, etc.  In order to 
remove this uncertainty one would like to find observables which are 
sensitive to the earliest times after the collision and are relatively 
unaffected by the later stages of plasma evolution.  One obvious 
candidate to consider is high-energy dilepton production since 
dileptons couple only electromagnetically to the plasma and therefore, 
in their high-energy spectra, carry information about plasma initial 
conditions.  To this end here we calculate the dependence of 
leading-order high-energy dilepton production on the assumed 
isotropization and thermalization time of the QGP using two simple 
models for early-time QGP evolution.

To begin the discussion we introduce two proper time scales: (1) the 
parton formation time, $\tau_0$, which is the time after which one can 
treat the partons generated from the nuclear collision by a distribution 
of on-shell partons; and (2) the isotropization time, $\tau_{\rm 
iso}$, which is the time when the system becomes isotropic in 
momentum-space. At RHIC energies $\tau_0 \sim 0.3$ fm/c and at LHC 
energies $\tau_0 \sim 0.1$ fm/c.  Immediately after the collision, the 
partons are produced from the incoming colliding nuclei at 
$\tau=\tau_0$, at which time the partonic momentum distributions can 
be assumed to be isotropic.\footnote{See 
Ref.~\cite{Jas:2007rw} for an alternative view of the early times 
after the initial collision wherein the authors find that the 
distribution may be prolate for longer than assumed here.  We postpone 
the study of the possibility of early-time prolate distributions to 
future work.} The subsequent rapid longitudinal expansion of the 
matter (along the beam line) causes it to become much colder in the 
longitudinal direction than in the transverse direction 
\cite{Baier:2000sb}. Longitudinal cooling occurs because initially the 
longitudinal expansion rate is larger than the parton interaction rate 
and, as a result, a local momentum-space anisotropy is induced with 
$\langle p^2_L\rangle\ll\langle p^2_T\rangle$ in the local rest frame. 
If the system is to return to an isotropic state it is necessary that 
at some later time the interaction rate overcomes the expansion rate 
with the system finally isotropizing and remaining isotropic for $\tau 
\geq \tau_{\rm iso}$. Once isotropy is achieved (and maintained by 
parton interactions) the use of hydrodynamic simulations can be 
justified.  We are therefore critically interested in the properties 
of the plasma around the isotropization time as these provide the 
relevant initial conditions for subsequent hydrodynamic evolution.

The study of anisotropic plasmas has received much interest recently 
due to the fact that a quark-gluon plasma which has a local 
momentum-space anisotropy, $2\langle p^2_L\rangle\neq\langle 
p^2_T\rangle$, is subject to the chromo-Weibel instability 
\cite{Strickland:2007fm,Mrowczynski:2000ed,Randrup:2003cw,%
Romatschke:2003ms,Arnold:2003rq,Romatschke:2004jh,Mrowczynski:2004kv,%
Rebhan:2004ur,Arnold:2005vb,Arnold:2004ti,Rebhan:2005re,%
Romatschke:2005pm,Schenke:2006xu,Schenke:2006fz,Manuel:2006hg,%
Bodeker:2007fw,Romatschke:2006nk,Romatschke:2006wg,Dumitru:2005gp,%
Dumitru:2006pz,Rebhan:2008uj}. The chromo-Weibel instability causes 
rapid growth of soft gauge fields which preferentially work to restore 
the isotropy of the quark-gluon plasma on time scales much shorter 
than the collisional time scale. However, most of the theoretical and 
numerical developments in describing the time-evolution of a QGP 
subject to the chromo-Weibel instability have been restricted to 
asymptotic energies at which perturbative resummations can be applied 
and hence the presence of the instability-driven isotropization at 
RHIC and LHC energies is not yet proven.  In addition, numerical 
studies of the chromo-Weibel instability in an 1-dimensionally 
expanding system show that there is a time delay before the effects of 
plasma instabilities become important to the system's dynamics 
\cite{Romatschke:2006wg,Rebhan:2008uj}. Future work will address these 
issues but until they become available there is a substantial amount 
of theoretical uncertainty in the QGP isotropization time, $\tau_0 
\leq \tau_{\rm iso} \lesssim$ 3 fm/c.

In the absence of a precise physical framework for describing the 
thermalization of the quark-gluon plasma and the associated time 
scales, one possible way to proceed is by studying the dependence of 
observables sensitive to the earliest times after the collision on the 
assumed plasma isotropization time by constructing simple space-time 
models. As mentioned above, one candidate observable is 
electromagnetic radiation such as high-energy photon and dilepton 
production\footnote{Hereafter, high-energy dileptons will refer to 
lepton pairs with pair transverse momentum ($p_T$) or invariant 
mass ($M$) greater than 1 GeV.} since these particles interact only 
electromagnetically and can escape from the strongly interacting 
medium created after the collision unhindered. Hence, they are perfect 
probes for studying the early-time dynamics of the system. In the case 
of high-energy medium photon production it is difficult for 
experimentalists to subtract the large backgrounds coming from $\pi^0$ 
decays from other sources of photons, making it hard to measure a 
clean high-energy medium photon production signal.\footnote{For a 
discussion of the effect of possible momentum-anisotropies on 
high-energy photon production see Ref. \cite{Schenke:2006yp}.} In the 
case of high-energy dileptons, the experimental situation is 
dramatically improved and it then becomes a question of making the 
necessary theoretical predictions to see how large the effect of a 
possible anisotropic pre-equilibrium phase would be on dilepton 
production.  

Phenomenological studies of the production of high-energy dileptons 
have shown that there are several important dilepton sources and it's 
necessary to include each of these depending on the kinematic region. 
For dilepton pair transverse momentum or invariant mass greater than 1 
GeV the most important sources of dilepton pairs are:  charm quark 
decays, initial state Drell-Yan scatterings,  jet-conversion, 
jet-fragmentation, and medium (thermal) production. For an extensive 
discussion of the various sources of the dilepton production in a 
heavy-ion collision, we refer the reader to 
\cite{Gale:2003iz,Turbide:2006zz} and references therein.  Note that 
in most previous phenomenological studies of photon and dilepton 
production it has been assumed that the system ``instantaneously'' 
thermalizes with $\tau_{\rm iso} = \tau_0$ and hence is locally 
isotropic throughout its evolution. 

In this paper, we extend our previous work \cite{Mauricio:2007vz} and 
concentrate on the impact of momentum-space anisotropies on the 
leading order medium dilepton production at large invariant mass and 
transverse momentum. We propose two simple phenomenological models for 
the time dependence of the plasma momentum-space anisotropy, 
$\xi=\frac{1}{2}\langle p^2_T\rangle/\langle p^2_L\rangle-1$, and hard 
momentum scale, $p_{\rm hard}$. In the first model we interpolate 
between 1+1 dimensional longitudinal free-streaming and 1+1 
dimensional ideal hydrodynamic expansion.  In the second model we 
incorporate the effect of momentum-space broadening due to hard-hard 
elastic scatterings in the pre-equilibrium dynamics.\footnote{We will 
assume that this elastic scattering rate is regulated in the infrared 
by an isotropic screening mass for simplicity.} In both models we 
introduce two parameters, $\tau_{\rm iso}$ and $\gamma$, with $\gamma$ 
setting the width of the transition from early-time pre-equilibrium 
dynamics to late-time equilibrated dynamics.  In the limit $\tau_{\rm 
iso} \rightarrow \tau_0$ our models reduce to ideal 1+1 dimensional 
hydrodynamical expansion and in the opposite limit $\tau_{\rm iso} 
\rightarrow \infty$ correspond to two different types of 
non-equilibrium plasma evolution:  free streaming expansion or 
collisionally-broadened expansion.

\begin{figure*}[t]
\begin{center}
\includegraphics[width=16.5cm]{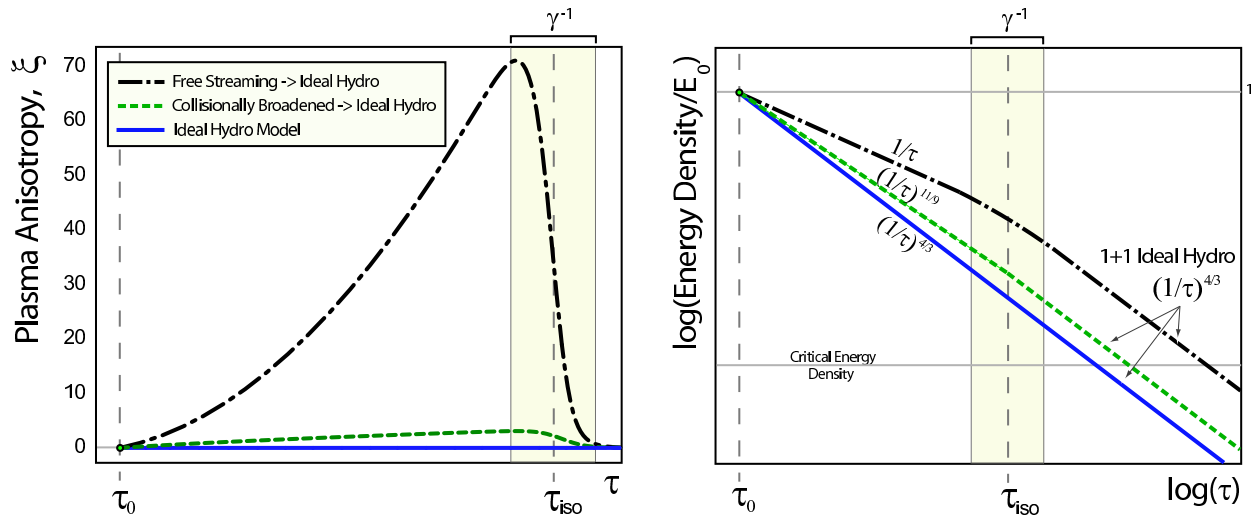}
\end{center}
\vspace{-2mm}
\caption{
Sketch of the time dependence of (left) the plasma anisotropy 
parameter, $\xi = \frac{1}{2} \langle p_T^2 \rangle/\langle p_L^2 \rangle -
1$, and (right) the normalized energy density, ${\cal E}/{\cal E}_0$, 
in the three different cases considered here.  Note that in the right
panel we only sketch the case of fixed initial conditions.  We also adjust
our models to constrain them to fixed final particle multiplicity.
}
\label{fig:modelSketch}
\end{figure*}

In Fig.~\ref{fig:modelSketch} we sketch the proper time dependence of 
the plasma anisotropy parameter, $\xi$, and energy density in the 
three cases considered below: free-streaming followed by ideal 
hydrodynamic expansion, collisionally broadened expansion followed by 
ideal hydrodynamic expansion, and ``instantaneous'' thermalization 
(ideal hydro throughout evolution). As can be seen from this figure 
for both the free streaming and collisionally-broadened cases with 
fixed initial conditions the energy density is always greater than 
that obtained by a system undergoing ideal hydrodynamic evolution and 
therefore one expects that, for fixed initial conditions, 
non-equilibrium effects can significantly enhance dilepton production. 
In order to calculate just how much the signal is enhanced requires 
detailed calculations which we present below.  As our first result we 
will show that for fixed initial conditions the addition of a 
pre-equilibrium phase at times $\tau < \tau_{\rm iso}$ enhances 
high-energy dilepton production by up to an order of magnitude.  

We then study the effect of constraining our space-time models so that 
the initial conditions vary in order to guarantee that the final pion 
multiplicity is independent of the assumed 
isotropization/thermalization time.  We show that allowing the initial 
conditions to vary in this way reduces the final effect of the 
possible anisotropic pre-equilibrium phase.  In order to quantify the 
effect we introduce a ratio called the ``dilepton enhancement'', 
$\phi$, which is the ratio of the dilepton yield obtained when one assumes 
a finite isotropization/thermalization time to that obtained when one 
assumes that the plasma ``instantaneously'' thermalizes at the 
formation time, $\tau_{\rm iso} = \tau_0$.  We show that, in our most 
realistic collisionally-broadened model, the dilepton enhancement can 
be up to 40\% at RHIC energies and 50\% at LHC energies if one 
assumes a isotropization/thermalization time of 2 fm/c.  We also show 
that as one varies the assumed isotropization/thermalization time that 
the dilepton enhancement, $\phi$, has a non-trivial dependence on pair 
transverse momentum which could allow experimental determination of 
$\tau_{\rm iso}$ if the experimental medium dilepton yields are 
obtained with high enough precision.

The work is organized as follows: In Sec. \ref{sec:dileptonrate} we 
calculate the dilepton production rate at leading order using an 
anisotropic phase space distribution. In Sec. 
\ref{sec:spacetimemodels} we construct models which interpolate 
between anisotropic and isotropic plasmas.  In Sec. \ref{sec:results} 
we present our final results on the dependence of dilepton production 
on the assumed plasma isotropization time and compare with other 
relevant sources of high-energy dileptons. Finally, we present our 
conclusions and give an outlook in the Sec. \ref{sec:conclusion}.


\section {Dilepton Rate from Kinetic Theory}
\label{sec:dileptonrate}

From relativistic kinetic theory, the dilepton production rate 
$dN^{l^+l^-}/d^4Xd^4P\equiv d R^{l^+l^-}/d^4P$ (i.e. the number of dileptons produced per space-time 
volume and four dimensional momentum-space volume) at leading order in
the electromagnetic coupling, $\alpha$, is 
given by \cite{Kapusta:1992uy,Dumitru:1993vz,Strickland:1994rf}:
\beq
\frac{d R^{l^+l^-}}{d^4P} = \int \frac{d^3{\bf p}_1}{(2\pi)^3}\,\frac{d^3{\bf p}_2}{(2\pi)^3}
			\,f_q({\bf p}_1)\,f_{\bar{q}}({\bf p}_2)\, \it{v}_{q\bar{q}}\,\sigma^{l^+l^-}_{q\bar{q}}\,
			\delta^{(4)}(P-p_1-p_2)
      \; ,
\label{eq:annihilation1}
\eeq
where $f_{q,{\bar q}}$ is the phase space distribution function of the medium quarks 
(anti-quarks), $\it{v}_{q\bar{q}}$ is the relative velocity between quark 
and anti-quark and $\sigma^{l^+l^-}_{q\bar{q}}$ is the total cross 
section
\beq
\sigma^{l^+l^-}_{q\bar{q}} = \frac{4\pi}{3} \frac{\alpha^2}{M^2} 
		\left(1 + \frac{2 m_l^2}{M^2}\right) 
		\left(1 - \frac{4 m_l^2}{M^2}\right)^{1/2} \; ,
\eeq
where $m_l$ is the lepton mass and $M$ is the center-of-mass energy. 
Since we will be considering high-energy dilepton pairs with 
center-of-mass energies much greater than the dilepton mass we can 
safely ignore the finite dilepton mass corrections and use simply 
$\sigma^{l^+l^-}_{q\bar{q}} = 4 \pi \alpha^2 / 3 M^2$. In addition, to 
very good approximation we can assume that the distribution function of 
quarks and anti-quarks is the same, $f_{\bar q}=f_q$.

In this work we will assume azimuthal symmetry of the matter in 
momentum-space so that the anisotropic quark/anti-quark phase 
distributions can be obtained from an arbitrary isotropic phase space 
distribution by squeezing ($\xi>0$) or stretching ($\xi<0$) along one 
direction in the momentum space, i.e. 
\beq
f_{q,{\bar q}}({\bf p},\xi,p_{\rm hard})=f^{iso}_{q,{\bar 
q}}(\sqrt{{\bf p^2}+\xi({\bf p\cdot \hat{n}}){\bf^2}},p_{\rm hard}) \; ,
\label{eq:distansatz}
\eeq
where $p_{\rm hard}$ is the hard momentum scale, $\hat{\bf n}$ is the 
direction of the anisotropy and $\xi >-1$ is a parameter that reflects 
the strength and type of anisotropy. In general, $p_{\rm hard}$ is 
related to the average momentum in the partonic distribution function. 
In isotropic equilibrium, where $\xi$=0, $p_{\rm hard}$ can be 
identified with the plasma temperature $T$.  To give another specific 
example, in the case of 1+1 dimensional free-streaming discussed in 
Sec.~\ref{freestreaming} $p_{\rm hard}$ is given by the initial 
``temperature'' $T_0$.

For general $\xi$ we split the delta function in 
Eq.~(\ref{eq:annihilation1}) such that we can perform the ${\bf p}_2$ 
integration:
\bqa
\frac{d R^{l^+l^-}}{d^4P} &=&
  \frac{5\alpha^2}{72\pi^5}\int\frac{d^3{\bf p}_1}{E_{{\bf p}_1}}\,  
  \frac{d^3{\bf p}_2}{E_{{\bf p}_2}}\,f_q({\bf p}_1,p_{\rm hard}, {\bf \xi})\,f_{\bar{q}}({\bf p}_2,p_{\rm hard}, {\bf \xi})\,\delta^{(4)}(P-p_1-p_2)      \nonumber \\
  &=& \frac{5\alpha^2}{72\pi^5}\int\frac{d^3{\bf p}_1}{E_{{\bf p}_1}E_{{\bf p}_2}}\, f_q({\bf p}_1,p_{\rm hard}, {\bf \xi})\, f_{\bar{q}}({\bf P}-{\bf p}_1,p_{\rm hard}, {\bf \xi})\,\delta (E-E_{{\bf p}_1}-E_{{\bf p}_2})
  \Biggr|_{{\bf p}_2={\bf P}-{\bf p}_1}.
\eqa
Choosing spherical coordinates with the anisotropy vector ${\bf 
\hat{n}}$ defining the $z$ axis, we can write:
\bqa
{\bf p}_1 &=&
p_1(\sin\theta_{p_1}\cos\phi_{p_1},\sin\theta_{p_1}\sin\phi_{p_1},\cos\theta_{p_1})\hspace{0.2cm},\nonumber
\\
{\bf P} &=&
P(\sin\theta_{P}\cos\phi_{P},\sin\theta_{P}\sin\phi_{P},\cos\theta_{P})\hspace{0.2cm}.
\eqa
It is then possible to reexpress the remaining delta function as:
\beq
\delta (E-E_{{\bf p}_1}-E_{{\bf p}_2})=2\,(E-p_1)\,\chi^{-1/2}\,\Theta(\chi)\sum_i^2\delta (\phi_i-\phi_{p_1})\hspace{0.2cm},
\label{delta}
\eeq
with $\chi\equiv\,4\,P^2\,p_1^2\,\sin^2\theta_P\,\sin^2\theta_{p_1}-(2p_1(E-P\cos\theta_P\cos\theta_{p_1})-M^2)^2$. The angles $\phi_i$ are defined as the solutions to the following transcendental equation:

\bqa
\cos\,(\phi_i-\phi_{p_1})=\frac{\,2\,p_1\,(E-P\cos\theta_P\cos\theta_{p_1})-M^2}{2\,P\,p_1\,\sin\theta_P\sin\theta_{p_1}}\hspace{0.2cm}.
\label{coseno}
\eqa
We point out that there are two solutions to Eq.~(\ref{coseno}) when 
$\chi>0$. After these substitutions and expanding out the phase space 
integrals, we obtain:
\bqa
\frac{d R^{l^+l^-}}{d^4P}&=&
\frac{5\alpha^2}{18\pi^5}\int_{-1}^1d(\cos\theta_{p_1})
\int_{a_+}^{a_-}\frac{dp_1}{\sqrt{\chi}}\,p_1\hspace{0.1cm}f_q\left({\sqrt{\bf p_1^2(1+\xi\cos^2\theta_{p_1})}},p_{\rm hard}\right)\nonumber \\
&&\times f_{\bar{q}}\left(\sqrt{{\bf(E-p_1)^2+\xi(p_1\cos\theta_{p_1}-P\cos\theta_P)^2}},p_{\rm hard}\right),
\label{scattering}
\eqa
with
\bqa
a_{\pm}&=&\frac{M^2}{2(E-P\cos (\theta_P\pm\theta_{p_1}))} \; .
\eqa
Note that when $\xi=0$, the limit of isotropic dilepton production is recovered trivially.
Also note that as $\xi$ increases we expect the differential dilepton rate to decrease 
since for fixed $p_{\rm hard}$ the increasing oblateness of the parton distribution functions
causes the effective parton density to decrease:
\bqa
n(\xi,p_{\rm hard}) &=& \int\,\frac{d^3p}{(2\pi)^3}\,f_q\,(\sqrt{{\bf p^2}+\,\xi\,({\bf p\cdot \hat{n}}){\bf^2}}\,,\,p_{\rm hard}) \; , \nonumber \\
	                  &=& \frac{n(\xi=0,p_{\rm hard})}{\sqrt{1+\xi}} \propto \frac{p_{\rm hard}^3}{\sqrt{1+\xi}} \; .
\eqa

\begin{figure*}[t]
\centerline{
\includegraphics[scale=0.48]{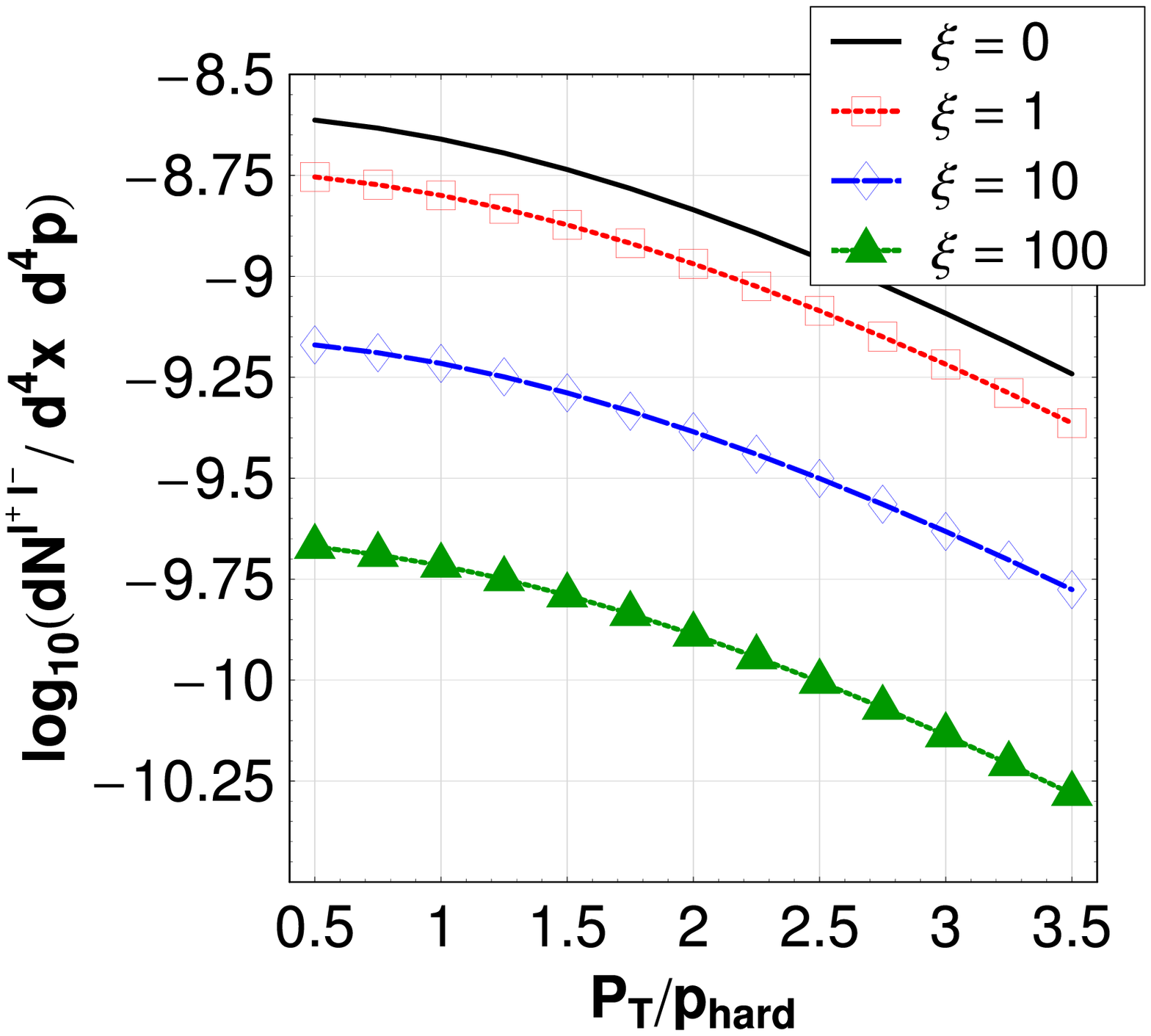}\hspace{5mm}
\includegraphics[scale=0.48]{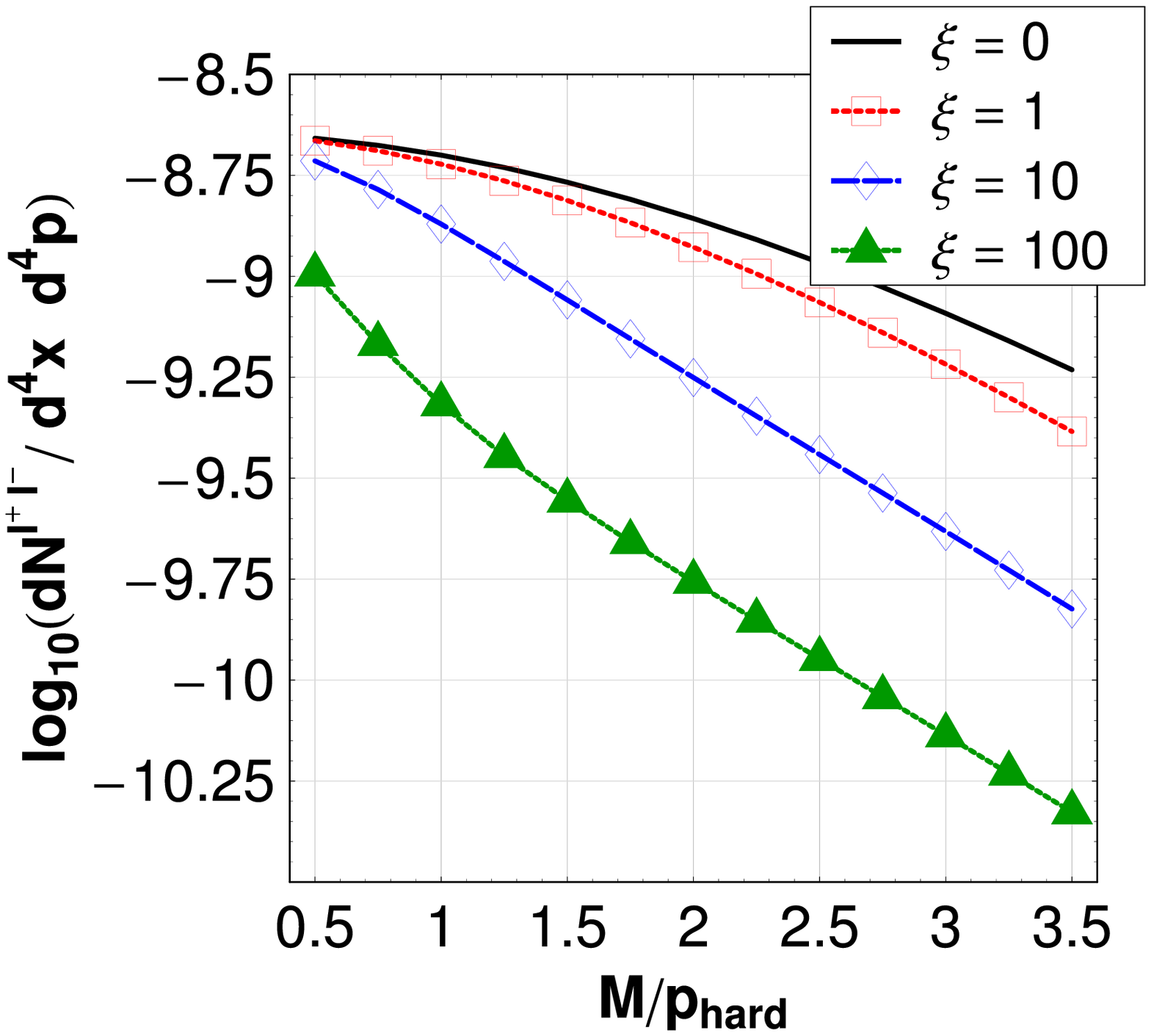}
}
\vspace{3mm}
\caption{
The differential dilepton rate as a function of transverse 
momentum (left) and invariant mass (right). For the invariant mass 
dependence (left) we fixed $p_T$= 3 GeV and for the transverse 
momentum dependence (right) we fixed $M$=3 GeV. In both cases $p_{\rm 
hard}$=1 GeV and rapidity $y$=0.
}
\label{diffrate}
\end{figure*}

In order to evaluate the anisotropic dilepton rate it is necessary to 
perform the remaining two integrations in Eq.~(\ref{scattering}) 
numerically. In Fig.~\ref{diffrate} we plot the resulting differential 
dilepton rate as a function of transverse momentum and invariant 
mass for $\xi \in \{0,1,10,100\}$. One can see the effect of 
increasing $\xi$ for fixed $p_{\rm hard}$, namely that the dilepton 
production rate decreases due, primarily, to the density effect 
mentioned above.

Knowing the rate, however, is not enough to make a phenomenological 
prediction for the expected dilepton yields. For this one must 
include the space-time dependence of $p_{\rm hard}$ and $\xi$ and 
then integrate over the space-time volume
\begin{subequations}
\begin{align}
  \frac{dN^{l^+l^-}}{dM^2dy}&=\pi R^2_T\int d^2p_T\int_{\tau_0}^{\tau_f}\int_{-\infty}^{\infty}\frac{dR^{l^+l^-}}{d^4P}\tau d\tau d\eta\hspace{0.2cm}, \label{Mspectrum}\\
        \frac{dN^{l^+l^-}}{d^2p_Tdy}&=\pi R^2_T\int dM^2\int_{\tau_0}^{\tau_f}\int_{-\infty}^{\infty}\frac{dR^{l^+l^-}}{d^4P}\tau d\tau d\eta\hspace{0.2cm}.\label{pTspectrum}
\end{align}
\label{spectrumeqs}
\end{subequations}
where $R_T\,=\,1.2\,A^{1/3}$ fm is the radius of the nucleus in the 
transverse plane. These expressions are evaluated in the center-of-mass
(CM) frame while the differential dilepton rate is calculated for 
the local rest frame (LR) of the emitting region. Then, the dilepton 
pair energy has to be understood as $E_{LR}=p_T\,\cosh\,(y-\eta)$ in 
the differential dilepton rate $dR_{\rm ann}/d^4P$. Additionally, in 
Eqs. (\ref{spectrumeqs}) we have assumed that there is only longitudinal expansion 
of the system. Since at early times the transverse expansion is small 
compared to the longitudinal expansion, one can ignore it. Some 
studies have suggested that the influence of the transverse expansion 
on the space-time evolution becomes phenomenologically important 
around 2.7 fm/c \cite{Ollitrault:2007du}, therefore, our approximation 
is valid for describing the early-time behaviour we are interested in. 
Substituting Eq.~(\ref{scattering}) into Eqs. (\ref{spectrumeqs}) we obtain the 
dilepton spectrum including the effect of a time-dependent momentum 
anisotropy. 

Note that we have not included the next-to-leading order (NLO) 
corrections to the dilepton rate due to the complexity of these 
contributions for finite $\xi$. These affect dilepton production for 
isotropic systems for $E/T\lesssim$ 1 \cite{Thoma:1997dk,%
Arnold:2002ja,Arleo:2004gn,Turbide:2006mc}. In the regions of phase 
space where there are large NLO corrections, we will apply $K$-factors 
to our results as indicated. These $K$-factors are determined by 
taking the ratio between NLO and LO calculation for an isotropic 
plasma, therefore, in this work we are implicitly assuming that the 
$K$-factor will be the same for an anisotropic plasma. 


\section {Space-Time Models}
\label{sec:spacetimemodels}

In this section we present two new models for 1+1 dimensional 
non-equilibrium time-evolution of the QGP and review the cases of 1+1 
dimensional free-streaming and 1+1 dimensional hydrodynamic expansion. 
In all cases considered below the number density will obey $n(\tau) 
\propto \left( \tau_{0}/\tau \right)$ in its asymptotic 
regions.\footnote{The interpolating models will only obey this 
relation outside of a region of order $\gamma^{-1} \tau_{\rm iso}$ 
around $\tau_{\rm iso}$ where the transition between different types 
of expansion takes place.  In the transition region $n$ will increase 
due to non-equilibrium effects as we discuss later in the text.} This 
results in all cases from the assumption that the total particle 
number is fixed while the size of the box containing the plasma is 
expanding at the speed of light in the longitudinal direction (1d 
expansion).

In each case below we will be required to specify a proper time 
dependence of the hard-momentum scale, $p_{\rm hard}$, and anisotropy 
parameter, $\xi$, which is consistent with this scaling for $\tau \ll 
\tau_{\rm iso}$ and $\tau \gg \tau_{\rm iso}$. Before proceeding, 
however, it is useful to note some general relations.  Firstly we 
remind the reader that the plasma anisotropy parameter is related 
to the average longitudinal and transverse momentum of the plasma 
partons via the relation
\beq
\xi=\frac{\langle p_T^2\rangle}{2\langle p_L^2\rangle} - 1 \; .
\label{anisoparam}
\eeq
Therefore, we can immediately see that for an isotropic plasma that $\xi=0$, 
and for an oblate plasma which has $\langle p_T^2\rangle > 2\langle 
p_L^2\rangle$ that $\xi>0$.

Secondly we note that given any anisotropic phase space distribution of the 
form specified in Eq.~(\ref{eq:distansatz}) the local energy density 
can be factorized via a change of variables to give
\begin{align}
\label{energy}
{\cal E}(p_{\rm hard},\xi) &= \int \frac{d^3{\bf p}}{(2\pi)^3}\hspace{0.1cm}p\hspace{0.1cm}
	f_{\rm iso}(\sqrt{{\bf p^2}+\xi({\bf p\cdot \hat{n}}{\bf^2})},p_{\rm hard}) \; , \\ 
&= {\cal E}_0(p_{\rm hard}) \, {\cal R}(\xi) \; ,\nonumber
\end{align}
where ${\cal E}_0$ is the initial local energy density deposited in 
the medium at $\tau_0$ and
\beq
{\cal R}(\xi) \equiv \frac{1}{2}\Biggl(\frac{1}{1+\xi}+\frac{\arctan\sqrt{\xi}}{\sqrt{\xi}} \Biggr) \; .
\label{calRdef}
\eeq
We note that $\lim_{\xi \rightarrow 0} {\cal R}(\xi) = 1$ and 
$\lim_{\xi \rightarrow \infty} {\cal R}(\xi) = 1/\sqrt{\xi}$.

\subsection{Asymptotic Limits of the Anisotropic Phase Space Distribution}
\label{subsec:asymptotic}

Before presenting our proposed interpolating models we review previous 
calculations for the free streaming and hydrodynamic expansion cases 
\cite{Kapusta:1992uy,Kampfer:1992bb,Kajantie:1986dh} and show how to determine our 
anisotropic phase space distribution function parameters, $p_{\rm 
hard}$ and $\xi$, in these two cases.


\subsubsection{1+1 Dimensional Ideal Hydrodynamical Expansion Limit}
\label{hydro}

We first consider the limiting case that $\tau_{\rm iso} = \tau_0$ so 
that the plasma is assumed to be ``instantaneously'' thermal and 
isotropic and undergoes ideal 1+1 dimensional hydrodynamical expansion 
throughout its evolution. In ideal hydrodynamical evolution using the 
boost-invariant 1+1 Bjorken model \cite{Bjorken:1982qr} we can identify 
$p_{\rm hard}$ with the temperature and the anisotropy parameter 
vanishes by assumption, $\xi=0$. Due to the fact that $\xi=0$ the 
distribution function for highly relativistic particles will depend 
only on the ratio between the energy and temperature, $f_{\rm 
hydro}(p,x) = f(E/T(\tau))$ with $E=(p_T^2+p_L^2)^{1/2}$.  In this 
case the number density, hard scale (temperature), energy density, and 
anisotropy parameter obey the following 
\begin{subequations}
\bqa
n(\tau) &=& n_0\; \left( \frac{\tau_{0}}{\tau} \right) \; , \label{hydroNeq} \\
p_{\rm hard}(\tau) &=& T(\tau) = T_0 \; \left( \frac{\tau_{0}}{\tau} \right)^{\frac{1}{3}} \; , \label{hydroTeq} \\
{\cal E}(\tau) &=& {\cal E}_0 \; \left( \frac{\tau_{0}}{\tau} \right)^{\frac{4}{3}} \; , \label{hydroEeq} \\
\xi(\tau) &=& 0 \; . \label{hydroXIeq}
\eqa
\end{subequations}

In order to obtain an analytic result for the differential dilepton 
rate which is applicable at high energies one can approximate the 
quark and anti-quark Fermi-Dirac distributions by Boltzmann 
distributions and integrate Eq.~(\ref{scattering}) analytically.  In 
this case it is also possible to perform the necessary integration of 
the rate over the plasma space-time evolution analytically 
\cite{Kapusta:1992uy, Kajantie:1986dh} to obtain:
\begin{subequations}
\begin{align}
\frac{dN^{l^+l^-}_{\rm hydro}}{dydM^2}&=\frac{5\alpha^2}{6\pi^2}\,\frac{1}{M^4}\,R^2_T\,T_0^6\,\tau_0^2
	\Biggl[ H\Biggl(\frac{M}{T_0}\Biggr)-H\Biggl(\frac{M}{T_c}\Biggr) \Biggr]\hspace{0.2cm}, \label{hydroM}\\
\frac{dN^{l^+l^-}_{\rm hydro}}{dM^2d^2p_Tdy}&=\frac{5\alpha^2}{24\pi^3}\,R^2_T\,\tau_0^2\Biggl(\frac{T_0}{m_T}\Biggr)^6
  \Biggl[ G\Biggl(\frac{m_T}{T_0}\Biggr)-G\Biggl(\frac{m_T}{T_c}\Biggr) \Biggr] \hspace{0.2cm},\label{hydroKT}
\end{align}
\end{subequations}
where $H(z)=z^2\,(8+z^2)\,K_0(z)+4\,z\,K_1(z)\,(4+z^2)$, 
$G(z)=z^3\,(8+z^2)\,K_3(z)$ and $m_T=\sqrt{M^2+p_T^2}$. As a check of 
our numerics we have verified that numerical integration of our dilepton 
rate given in Eq.~(\ref{scattering}) over space-time via 
Eqs.~(\ref{spectrumeqs}) reproduces this analytic result in the limit 
$\tau_{\rm iso} \rightarrow \tau_0$ and $\xi=0$.


\subsubsection{1+1 Dimensional Free Streaming Limit}
\label{freestreaming}

As another limiting case we can assume instead that our 1+1 
dimensional expanding plasma is non-interacting.  If this were true 
then the system would simply undergo 1+1 dimensional free-streaming 
expansion \cite{Kapusta:1992uy,Kampfer:1992bb}. Since, in this case, 
the system would never become truly thermal or isotropic this 
corresponds to taking the opposite limit from the one we took in the 
previous subsection, namely we will now take the limit $\tau_{\rm iso} 
\rightarrow \infty$.
 
In the free streaming case, the distribution function is a solution of 
the collisionless Boltzmann equation
\bqa
p\,\cdot\,\partial_x\,f_{\rm f.s.}(p,x)=0 \;,
\label{eq:freestreamb}
\eqa
where the subscript ${\rm f.s.}$ indicates that this is
the free-steaming solution.
In this work we will also assume that the distribution
function is isotropic at the formation time, $\tau=\tau_0$.
\begin{equation}
f_{\rm f.s.}(p,x)\Biggr|_{\bf \tau=\tau_0}=f\Biggl(\frac{\sqrt{p_T^2+p_L^2}}{p_{\rm hard}}\Biggr) \; ,
\end{equation}
where $p_T$ is the transverse momentum, $p_L$ is the longitudinal 
momentum and $p_{\rm hard}$ is the hard momentum scale at $\tau_0$. 
The typical hard momentum scale of particles undergoing 
1+1 dimensional free streaming expansion is constant in time.  In the 
case of indefinite free-streaming expansion the system never reaches 
thermal equilibrium and so the system strictly cannot have a 
temperature associated with it; however, since our assumed 
distribution function is isotropic at $\tau=\tau_0$, we can identify 
the initial ``temperature'' of the system, $T_0$, with the hard 
momentum scale $p_{\rm hard}$ when comparing hydrodynamic and free 
streaming expansion.

Eq.~(\ref{eq:freestreamb}) has a family of solutions which are
boost invariant along the $z$ (beam) axis
\bqa
f_{\rm f.s.}(p,x)\,=\,f\,(\,p_T\,,p_L\,t\,-\,E\,z\,)\hspace{0.2cm}.
\eqa
Therefore, the functional dependence of the distribution function for 
the free streaming case is of the form
\bqa
f_{\rm f.s.}(p,x)=\,f\Biggl(\frac{\sqrt{p_T^2+(p_L t-E z)^2/\tau_0^2}}{T_0}\Biggr)\hspace{0.2cm}.
\label{eq:fssol}
\eqa

This distribution function can be simplified if we change to co-moving coordinates:
\begin{subequations}
\label{comovcoord}
 \begin{align}
p_L\,&=\,p_T\,\sinh y \hspace{0.2cm},\hspace{0.5cm}E\,=\,p_T\,\cosh y \hspace{0.2cm},\\
z\,&=\,\tau\,\sinh \eta \hspace{0.2cm},\hspace{0.85cm}t\,=\,\tau\,\cosh \eta \hspace{0.2cm},
\end{align}
\end{subequations}
where, as usual, $y$ is the momentum-space rapidity, $\tau$ is the proper time, and 
$\eta$ is the space-time rapidity. In terms of these variables one obtains
\bqa
\label{fsfunction}
f_{\rm f.s.}(p,x) = 
  f\Biggr(\frac{p_T}{T_0}\sqrt{1+\frac{\tau^2}{\tau_0^2}\sinh^2\,(y-\eta)}\,\Biggr) \; .
\label{eq:fssol2}
\eqa
Note that in the case of indefinite free-streaming at late times the 
quark and anti-quark longitudinal momentum are highly red-shifted 
reducing late time emission of high-energy dilepton pairs.
 
As written in Eq.~(\ref{anisoparam}) the anisotropy parameter is 
related with the average transverse and longitudinal momenta of the 
partons.  The average momentum-squared values appearing there are 
defined in the standard way:
\beq
<p_{T,L}^2> \, \equiv \, \frac{ \int \! d^3{\bf p} \, p_{T,L}^2 \, f(p,x) }{
                          \int \! d^3{\bf p} \, f(p,x) } \; .
\label{eq:ptlavg}
\eeq
Using the 1+1 dimensional free streaming distribution given in 
Eq.~(\ref{eq:fssol2}) and transforming to co-moving coordinates 
defined in (\ref{comovcoord}) so that $d^3{\bf p} \rightarrow p_T^2 
\cosh y \, dp_T \, dy$ we obtain
\begin{subequations}
 \begin{align}
  \langle p_T^2 \rangle_{\rm f.s.} &\propto 2 \, T_0^2 \; ,\\
  \langle p_L^2 \rangle_{\rm f.s.} &\propto T_0^2 \frac{\tau_0^2}{\tau^2} \; .
 \end{align}
\end{subequations}
Inserting these expressions into the general expression for $\xi$ 
given in Eq.~(\ref{anisoparam}) one obtains $\xi_{f.s.}(\tau) = 
\tau^2/\tau_0^2-1$. With this in hand we can also determine 
proper time dependence of the energy density in the free-streaming 
case by substituting this expression for $\xi$ into 
Eq.~(\ref{energy}), ${\cal E}_{\rm f.s.}(\tau) = {\cal E}_0 \, {\cal 
R}(\xi_{f.s.}(\tau))$. At early times one must use the full expression 
given by Eq.~(\ref{energy}); however, at late times one can expand 
this result to obtain ${\cal E}_{\rm f.s.}(\tau) \propto \tau_0/\tau$ 
as expected for a 1+1 free streaming plasma \cite{Baym:1984np}.

Summarizing, one finds in the 1+1 free streaming case that
in the limit $\tau \gg \tau_0$:
\begin{subequations}
\bqa
n(\tau) &=& n_0 \; \left( \frac{\tau_{0}}{\tau} \right) \; , \label{fsNeq} \\
p_{\rm hard}(\tau) &=& p_{\rm hard}(\tau=0) = T_0 \; , \label{fsTeq} \\
{\cal E}(\tau) &=& {\cal E}_0 \; \left( \frac{\tau_{0}}{\tau} \right) \; , \label{fsEeq} \\
\xi(\tau) &=& \frac{\tau^2}{\tau_0^2} - 1 \; . \label{fsXIeq}
\eqa
\label{fslimit}
\end{subequations}

With the distribution function given by Eq.~(\ref{fsfunction}), the 
dilepton spectrum can be calculated. As a function of the invariant mass 
$M$, one obtains
\begin{align}
\frac{dN^{l^+l^-}_{\rm f.s.}}{dydM^2}&=\frac{5\alpha^2}{72\pi^3}R^2_T\,M^2\,\tau_0^2\int x_1\hspace{0.1cm}x_2\hspace{0.1cm}dx_1\hspace{0.1cm}dx_2\hspace{0.1cm}dy_1\hspace{0.1cm}dy_2\hspace{0.1cm}d(\tau/\tau_0)^2\nonumber\\
&\times [(x_1x_2)^2-(x_1x_2\cosh\,(y_1-y_2)-1/2)^2\,]^{-1/2}\nonumber\\
&\times f^q_{\rm f.s.}\Biggl(\frac{M}{T_0}x_1\sqrt{1+\Bigl(\frac{\tau}{\tau_0}\Bigr)^2\sinh^2 y_1}\hspace{0.1cm}\Biggr)\hspace{0.1cm}f^{\bar{q}}_{\rm f.s.}\Biggl(\frac{M}{T_0}x_2\sqrt{1+\Bigl(\frac{\tau}{\tau_0}\Bigr)^2\text{sinh}^2\, y_2}\hspace{0.1cm}\Biggr)\hspace{0.2cm}.\label{freeM}
\end{align}
In the last expression, the integration is over all $x_i$ from 0 to 
$+\infty$ and over all $y_i$ from $-\infty$ to $\infty$ subject to the 
constraint
\beq
\frac{1}{\cosh (y_1-y_2)+1}\leqslant 2x_1x_2\leqslant\frac{1}{\cosh (y_1-y_2)-1}\hspace{0.2cm}.
\nonumber
\eeq

As a function of the transverse momentum, $p_T$, the dilepton 
production using free streaming case we obtain\footnote{In 
the original article by Kapusta et. al \cite{Kapusta:1992uy}, the 
calculation of $dN^{l^+l^-}/dydM^2d^2p_T$ was not presented.}
\begin{align}
\frac{dN^{l^+l^-}_{\rm f.s.}}{dM^2d^2p_Tdy}&=\frac{5\alpha^2}{36\pi^4}R^2_T\tau_0^2\int_{x_+}^{x_-} x\hspace{0.1cm}dx\hspace{0.1cm}dy_1\hspace{0.1cm}dy_2\hspace{0.1cm}d(\tau/\tau_0)^2\hspace{0.1cm}f^q_{\rm f.s.}\Biggl(\frac{x\hspace{0.1cm}M}{T_0}\sqrt{1+\Bigl(\frac{\tau}{\tau_0}\Bigr)^2\sinh^2y_1}\Biggr)\nonumber\\
\times 
 f^{\bar{q}}_{\rm f.s.}\Biggl(\frac{M}{T_0}&\Bigl(\Bigl(\frac{m_T}{M}\Bigr)^2+x^2-2\frac{m_T}{M}x\cosh (y_1-y_2)+\Bigl(\frac{\tau}{\tau_0}\Bigr)^2\Bigl(\frac{m_T}{M}
\text{sinh}\, y_2-x\sinh y_1\Bigr)^2\Bigr)^{1/2}\Biggr)\nonumber\\
&\times \Bigl\{\Bigl(\frac{p_T}{M}\, x\Bigr)^2-\Bigl(\frac{m_T}{M}\, x \,\cosh (y_1-y_2)-\frac{1}{2}\Bigr)^2\Bigr\}^{-1/2}\hspace{0.2cm},\label{freekT}
\end{align}
with
\begin{equation}
x_\pm=\frac{M}{2\,(m_T\,\cosh\,(y_1-y_2)\,\pm\,p_T)} \; .
\end{equation}
We have verified that using the expressions listed in Eq.~(\ref{fslimit}) 
our direct numerical integration of the rate given in 
Eq.~(\ref{scattering}) over space-time via Eqs.~(\ref{spectrumeqs}) 
reproduces this analytic result in the free-streaming limit.

We note in closing that as a solution of the collisionless 
(non-interacting) Boltzmann equation, the free-streaming case can be 
taken as an upper bound on the magnitude of the plasma anisotropy 
parameter since for fixed $\langle p_T^2 \rangle$ (no transverse 
expansion/contraction) $\xi$ cannot be larger than the free-streaming 
value by causality.


\subsection{Momentum-space Broadening in a 1+1 Dimensionally Expanding Plasma}
\label{subsec:momentumbroadening}

In the previous two subsections we presented details of the limiting 
cases for 1+1 dimensional plasma evolution:  1+1 ideal hydrodynamic 
expansion and 1+1 dimensional free streaming, with the former arising 
if there is rapid thermalization of the plasma and the latter arising 
if the plasma has no interactions.  We would now like to extend these 
models to include the possibility of momentum-space broadening of the 
plasma partons due to interactions (hard and soft).  This can be accomplished 
mathematically by generalizing our expression for $\xi(\tau)$ to 
\beq
\xi(\tau,\delta) = \left( \frac{\tau}{\tau_0} \right)^\delta - 1 \; .
\label{broadenedxi}
\eeq
In the limit $\delta \rightarrow 0$, $\xi \rightarrow 0$ and one 
recovers the 1+1 hydrodynamical expansion limit and in the limit 
$\delta \rightarrow 2$ one recovers the 1+1 dimensional free streaming 
limit, $\xi \rightarrow \xi_{\rm f.s.}$ For general $\delta$ between 
these limits one obtains the proper time dependence of the energy 
density and temperature by substituting (\ref{broadenedxi}) into the 
general expression for the factorized energy density (\ref{energy}) to obtain 
${\cal E}(\tau,\delta) = {\cal E}_0 \, {\cal R}(\xi(\tau,\delta))$. 
In the limit $\tau \gg \tau_0$ this gives the following scaling 
relations for the number density, energy density, and hard momentum 
scale
\begin{subequations}
\label{deltageneral}
\bqa
n(\tau) &=& n_0 \, \left( \frac{\tau_0}{\tau} \right) \; , \label{deltaNeq} \\
p_{\rm hard}(\tau) &=& T_0 \, \left( \frac{\tau_0}{\tau} \right)^{\left(1-\delta/2\right)/3}  \; , \label{deltaTeq} \\
{\cal E}(\tau) &=& {\cal E}_0 \, \left( \frac{\tau_0}{\tau} \right)^{4\left(1-\delta/8\right)/3} \; . \label{deltaEeq}
\eqa
\end{subequations}
Different values of $\delta$ arise dynamically from the different 
processes contributing to parton isotropization.  Below we list the 
values of $\delta$ resulting from processes which are relevant during 
the earliest times after the initial nuclear impact.

\subsubsection{Collisional Broadening via Elastic 2$\leftrightarrow$2 collisions}

In the original version of the bottom up scenario \cite{Baier:2000sb}, 
it was shown that, even at early times after the nuclear impact, 
elastic collisions between the liberated partons will cause a 
broadening of the longitudinal momentum of the particles compared to 
the non-interacting, free-streaming case. During the first stage of 
the bottom-up scenario, when $1\ll Q_s\tau\ll\alpha_s^{3/2}$, the initial 
hard gluons have typical momentum of order $Q_s$ and occupation number 
of order $1/\alpha_s$. Due to the fact that the system is expanding at 
the speed of light in the longitudinal direction $N_g \sim 
Q_s^3/(\alpha_s Q_s\tau)$. If there were no interactions this expansion 
would be equivalent to 1+1 free streaming and the longitudinal 
momentum $p_L$ would scale like $1/\tau$. However, when elastic 
$2\leftrightarrow 2$ collisions of hard gluons are taken into account 
\cite{Baier:2000sb}, the ratio between the longitudinal momentum $p_L$ 
and the typical transverse momentum of a hard particle $p_T$ decreases 
as:
\begin{equation}
\label{ptbroadbottom}
\frac{\langle p_L^2 \rangle}{\langle p_T^2 \rangle} \propto (Q_s\tau)^{-2/3} \; .
\end{equation}
Assuming, as before, isotropy at the formation time, 
$\tau_0=Q_s^{-1}$, this implies that for a collisionally-broadened 
plasma $\delta=2/3$.  Note that, as obtained in 
Ref~\cite{Baier:2000sb}, the derivation of this result makes an 
implicit assumption that the elastic cross-section is screened at long 
distances by an isotropic real-valued Debye mass.  This is not 
guaranteed in an anisotropic plasma as the Debye mass can be become 
complex due to the chromo-Weibel instability \cite{Romatschke:2003ms}. 
However, at times short compared to the time scale where plasma 
instabilities become important we expect the isotropic result to hold 
to good approximation.

\subsubsection{Effect of Plasma Instabilities}

Plasma instabilities affect the first stage of bottom-up scenario 
\cite{Arnold:2003rq}. These instabilities are characterized by the 
growing of chromo-electric and -magnetic fields $E^{a}$ and 
$B^{a}$. These fields bend the particles and how much bending occurs 
will depend on the amplitude and domain size of the induced 
chromofields. Currently, the precise parametric relations between the 
amount of plasma anisotropy and amplitude and domain size of the 
chromofields are not known from first principles.  There are three 
possibilities for how the chromo-Weibel instability will affect 
isotropization of a QGP proposed in the literature 
\cite{Bodeker:2005nv, Arnold:2005qs, Arnold:2007cg}:
\begin{equation}
\label{ptbroadproposals1}
\frac{\langle p_L^2 \rangle}{\langle p_T^2 \rangle}\sim(Q_s\tau)^{-\frac{1}{2}\bigl(\frac{1}{1+\nu}\bigl)} \; ,
\end{equation}
where
\beq
\label{ptbroadproposals2}
\nu=\left\{ \begin{aligned}
0 \hspace{0.2cm}&\text{Ref.\cite{Bodeker:2005nv} \; ,}\\
1 \hspace{0.2cm}&\text{Ref.\cite{Arnold:2005qs} \; ,}\\
2 \hspace{0.2cm}&\text{Nielsen-Olesen limit, Ref.\cite{Arnold:2007cg}} \;.
          \end{aligned}
	  \right.
\eeq
These results correspond to $\delta=1/2$, $\delta=1/4$, and 
$\delta=1/6$, respectively.

\subsubsection{Summary and Discussion}

Summarizing, the coefficient $\delta$ takes on the following values
\beq
\label{anisomodels}
\delta=\left\{ 
\begin{aligned}
2 \hspace{0.2cm}&\text{Free streaming expansion} \; ,\\
2/3 \hspace{0.2cm}&\text{Collisional-Broadening, Ref.\cite{Baier:2000sb}\; ,}\\
1/2 \hspace{0.2cm}&\text{Ref.\cite{Bodeker:2005nv}}\; , \\
1/4 \hspace{0.2cm}&\text{Ref.\cite{Arnold:2005qs}\; ,}\\
1/6 \hspace{0.2cm}&\text{Nielsen-Olesen limit, Ref.\cite{Arnold:2007cg}}\; ,\\
0\hspace{0.2cm}&\text{Hydrodynamic expansion} \; .
\end{aligned}
\right.
\eeq
The exponents in Eq.~(\ref{anisomodels}) are direct consequence of the 
relation between the anisotropy parameter $\xi$ and the longitudinal 
and transverse momentum given in Eq.~(\ref{anisoparam}). The exponent 
$\delta$ indicates which kind of broadening we are considering. Notice 
that $\delta$=2 (0) reproduces the behaviour of free streaming 
(hydrodynamic) expansion. 

\begin{figure*}[t]
\includegraphics[width=10cm]{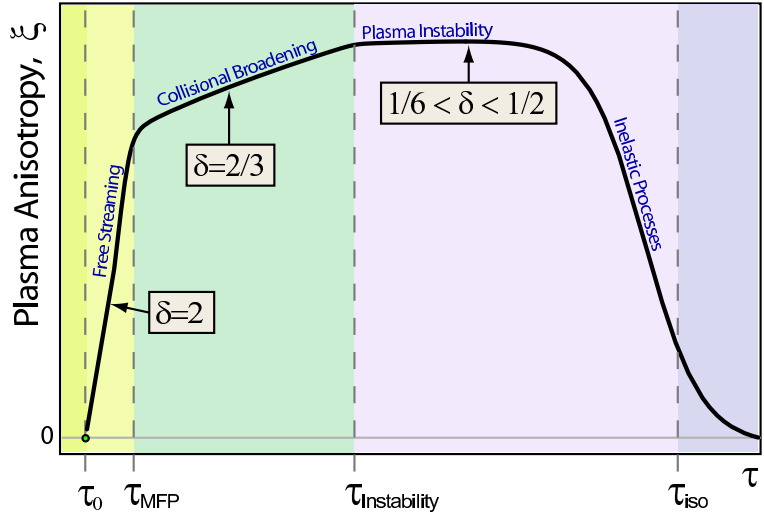}
\vspace{-2mm}
\caption{
Sketch of the time dependence the plasma anisotropy indicating the 
various time-scales and processes taking place.  Here $\tau_{\rm MFP}$ 
is the time between elastic collisions (mean-free time) and $\tau_{\rm 
Instability}$ is the time at which plasma-instability induced soft 
modes have grown large enough to affect hard particle dynamics.
}
\label{fig:deltaSketch}
\end{figure*}

In Fig.~\ref{fig:deltaSketch} we sketch the time-dependence of the 
plasma anisotropy parameter indicating the time scales at which the 
various processes become important.  At times shorter than the mean 
time between successive elastic scatterings, $\tau_{\rm MFP}$, the 
system will undergo 1+1 dimensional free streaming with $\delta=2$. 
For times long compared to $\tau_{\rm MFP}$ but short compared to 
$\tau_{\rm Instability}$ the plasma anisotropy will grow with the 
collisionally-broadened exponent of $\delta=2/3$. Here $\tau_{\rm 
Instability}$ is the time at which instability-induced soft gauge 
fields begin to influence the hard-particles' motion. When $\tau_{\rm 
Instability} < \tau < \tau_{\rm iso}$ the plasma anisotropy grows with 
the slower exponent of $\delta = 1/6 \ldots 1/2$ due to the bending of 
particle trajectories in the induced soft-field background.  At times 
large compared to $\tau_{\rm Instability}$ inelastic processes are 
expected to drive the system back to isotropy \cite{Baier:2000sb}.
We note here that for small $\xi$ and realistic couplings it has been 
shown \cite{Schenke:2006xu} that one cannot ignore the effect of 
collisional-broadening of the distribution functions and that this may
completely eliminate unstable modes from the spectrum.

Based on such a sketch one could try to construct a detailed model 
which includes all of the various time scales and study the dependence 
of the process under consideration on each.  However, due to the 
current theoretical uncertainties in each of these time scales and 
their dependences on experimental conditions we choose to use a 
simpler approach in which we will construct two phenomenological 
models which smoothly interpolate the coefficient $\delta$:
\begin{align*}
\text{Free streaming interpolating model :} & \;\; 2 \geq \delta \geq 0 \; , \\ 
\text{Collisionally-broadened interpolating model :} & \;\; \frac{2}{3} \geq \delta \geq 0 \; . 
\end{align*}
In both models we introduce a transition width, $\gamma^{-1}$, which 
governs the smoothness of the transition from the initial value of 
$\delta \in \{2,2/3\}$ to $\delta=0$ at $\tau \sim \tau_{\rm iso}$. 
The free streaming interpolating model will serve as an upper-bound on 
the possible effect of early time momentum-space anisotropies while 
the collisionally-broadened interpolating model should provide a more 
realistic estimate of the effect due to the lower anisotropies 
generated.  This will help us gauge our theoretical uncertainties. 
Note that by using such a smooth interpolation one can achieve a 
reasonable phenomenological description of the transition from 
non-equilibrium to equilibrium dynamics which should hopefully capture 
the essence of the physics. In the next section we will give mathematical 
definitions for these two models.

\begin{figure*}[t]
\begin{center}
\includegraphics[width=17.7cm]{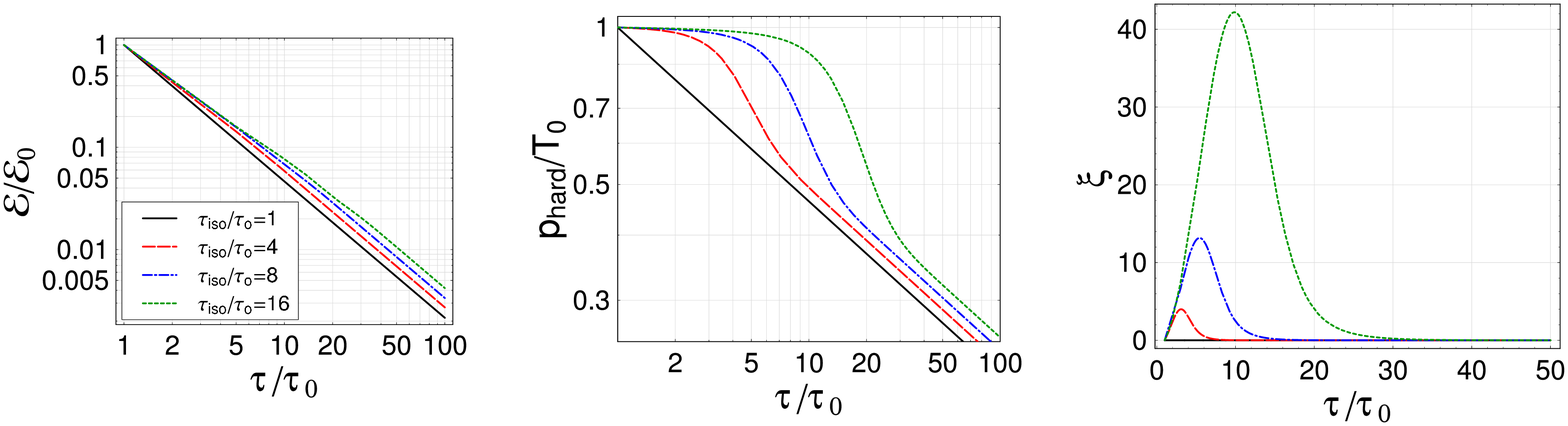}
\vspace{2mm}
\includegraphics[width=17.7cm]{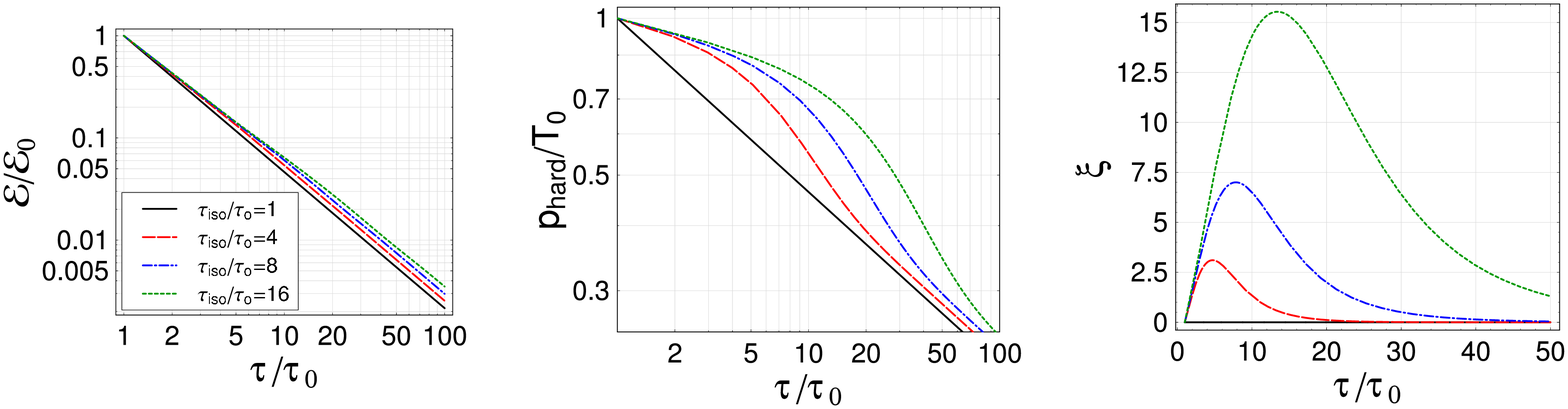}
\end{center}
\caption{Temporal evolution using our fixed initial condition free-streaming interpolating model 
($\delta = 2$) for the energy density (left column), hard momentum 
scale (middle column), and anisotropy parameter (right column) for 
four different isotropization times $\tau_{\rm iso} \in \{1,4,6,18\} 
\, \tau_0$. The transition width is taken to be (top row) $\gamma = 2$ 
and (bottom row) $\gamma = 0.5$.  To convert to physical scales use
$\tau_0 \sim 0.3$ fm/c for RHIC and $\tau_0 \sim 0.1$ fm/c for LHC.
}
\label{fig:modelPlotFS}
\end{figure*}


\subsection{Space-Time Interpolating Models with Fixed Initial Conditions}
\label{subsec:model}

In order to construct our interpolating models, the parameter 
$\delta$ should be a function of proper time. To 
accomplish this, we introduce a smeared step function
\beq
\lambda(\tau,\tau_{\rm iso},\gamma) \equiv \frac{1}{2} \left({\rm 
tanh}\left[\frac{\gamma (\tau-\tau_{\rm iso})}{\tau_{\rm iso}} \right]+1\right) \; ,
\eeq
where $\gamma^{-1}$ sets the width of the transition between 
non-equilibrium and hydrodynamical evolution in units of $\tau_{\rm 
iso}$.\footnote{Note that compared to Ref.~\cite{Mauricio:2007vz} we 
have modified our definition of $\gamma$ so that it now measures the 
width in units of $\tau_{\rm iso}$ instead of $\tau_0$. This results 
in time-dependence of modeled quantities not experiencing unphysical 
``dips'' which can occur for large values of $\gamma$ in our previous 
interpolating model \cite{AndiPersonal}.} In the limit when $\tau \ll 
\tau_{\rm iso}$, we have $\lambda \rightarrow 0$ and 
when $\tau \gg \tau_{\rm iso}$ we have $\lambda \rightarrow 1$.

Physically, the energy density ${\cal E}$ should be continuous
as we change from the initial non-equilibrium value of $\delta$
to the final isotropic $\delta=0$ value appropriate for ideal hydrodynamic
expansion.  Once the energy
density is specified this immediately gives us the time dependence
of the hard momentum scale.  We find that for general $\delta$ this
can be accomplished with the following model
\begin{subequations}
\label{eq:modelEQs}
\begin{align}
\xi(\tau,\delta) &= \left(\tau/\tau_0\right)^{\delta(1-\lambda(\tau))} - 1 \; , \\
{\cal E}(\tau) &= {\cal E}_0 \; {\cal R}\left(\xi\right) \; \bar{\cal U}^{4/3}(\tau) \; , \\
p_{\rm hard}(\tau) &= T_0 \; \bar{\cal U}^{1/3}(\tau) \; ,  
\label{pdependence}
\end{align}
\end{subequations}
with ${\cal R}(\xi)$ defined in Eq.~(\ref{calRdef}) and for fixed initial conditions
\begin{subequations}
\bqa
{\cal U}(\tau) &\equiv& \left[{\cal R}\!\left(\left(\tau_{\rm iso}/\tau_0\right)^\delta- 
1\right)\right]^{3\lambda(\tau)/4} \left(\frac{\tau_{\rm iso}}{\tau}\right)^{1 - \delta\left(1-\lambda(\tau)\right)/2} \; ,  \\
\bar{\cal U}(\tau) &\equiv& {\cal U}(\tau) \, / \, {\cal U}(\tau_0) \; .
\eqa
\label{Udeff}
\end{subequations}
The power of ${\cal R}$ in ${\cal U}$ keeps the energy density 
continuous at $\tau = \tau_{\rm iso}$ for all $\gamma$. In the 
following subsections we will briefly discuss the two interpolating 
models we consider in this work.

\subsubsection{Free streaming interpolating model}

Using Eq.~(\ref{eq:modelEQs}) we can obtain a model which interpolates 
between early-time 1+1 dimensional longitudinal free streaming and 
late-time 1+1 dimensional ideal hydrodynamic expansion by choosing 
$\delta=2$.  With this choice and in the limit $\tau \ll 
\tau_{\rm iso}$, we have $\lambda \rightarrow 0$ and the system 
undergoes 1+1 dimensional free streaming. When $\tau \gg \tau_{\rm iso}$ then $\lambda 
\rightarrow 1$ and the system is expanding hydrodynamically.  In the 
limit $\gamma \rightarrow \infty$, $\lambda \rightarrow 
\Theta(\tau-\tau_{\rm iso})$, the system makes a theta function 
transition from free streaming to hydrodynamical evolution with the energy 
density being continuous during this transition by construction. In 
Fig.~\ref{fig:modelPlotFS} we plot the time-dependence of ${\cal 
E}$, $p_{\rm hard}$, and $\xi$ assuming (top) $\gamma=2$ and (bottom) 
$\gamma=0.5$ for different values of $\tau_{\rm iso}$.  As can be 
seen from this figure for fixed initial conditions during the period 
of free-streaming evolution  the system always has a higher effective 
temperature ($p_{\rm hard}$) than would be obtained by a system which 
undergoes only hydrodynamic expansion from the formation time.  As we 
will show in the results section, for fixed initial conditions, this 
results in a sizable enhancement in high-energy dilepton production.

\begin{figure*}[t]
\includegraphics[width=17.7cm]{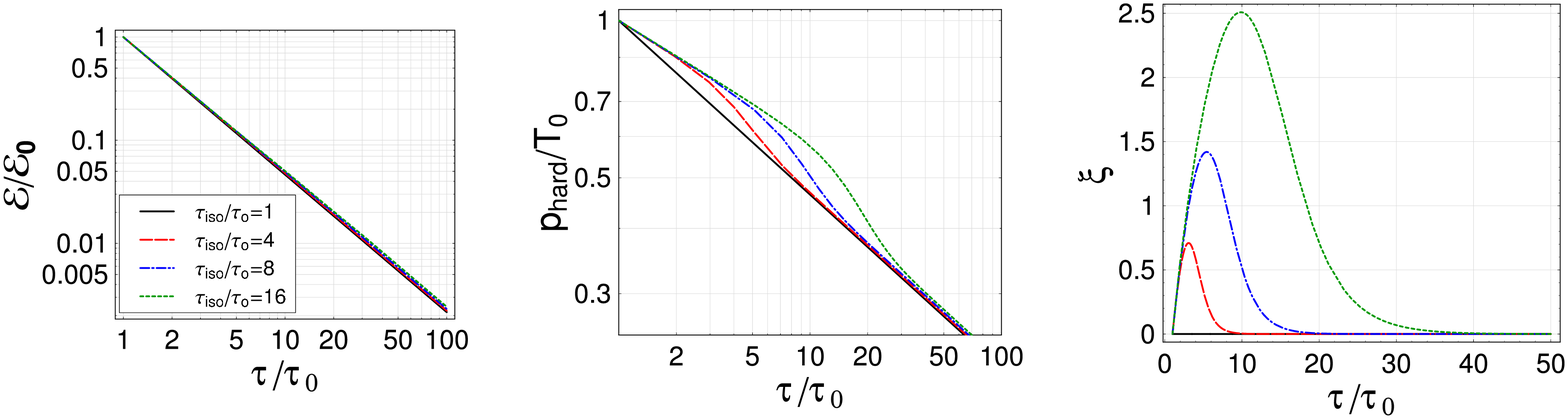}

\vspace{2mm}
\includegraphics[width=17.7cm]{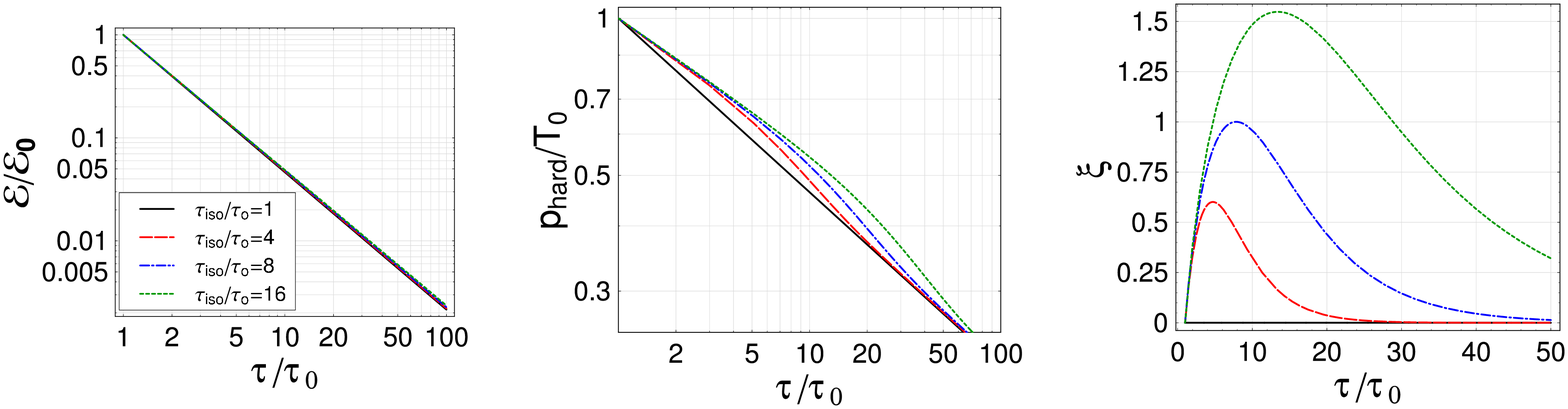}
\caption{Temporal evolution using our fixed initial condition collisionally-broadened interpolating model 
($\delta = 2/3$) for the energy density (left column), hard momentum 
scale (middle column), and anisotropy parameter (right column) for 
four different isotropization times $\tau_{\rm iso} \in \{1,4,6,18\} 
\, \tau_0$. The transition width is taken to be (top row) $\gamma = 2$ 
and (bottom row) $\gamma = 0.5$.  To convert to physical scales use
$\tau_0 \sim 0.3$ fm/c for RHIC and $\tau_0 \sim 0.1$ fm/c for LHC.}
\label{fig:modelPlotCB}
\end{figure*}

\subsubsection{Collisionally-broadened interpolating model}

Similarly using Eq.~(\ref{eq:modelEQs}) we can obtain a model which 
interpolates between early-time 1+1 dimensional 
collisionally-broadened expansion and late-time 1+1 dimensional ideal 
hydrodynamic expansion by choosing $\delta=2/3$.  In 
Fig.~\ref{fig:modelPlotCB} we plot the time-dependence of ${\cal E}$, 
$p_{\rm hard}$, and $\xi$ assuming (top) $\gamma=2$ and (bottom) 
$\gamma=0.5$ for different values of $\tau_{\rm iso}$.  As in the 
free-streaming interpolating model for fixed initial conditions at 
early times a collisionally-broadened system always has a higher 
effective temperature ($p_{\rm hard}$) than would be obtained by a 
system which undergoes only hydrodynamic expansion from the formation 
time.  As we will show in the results section, for fixed initial 
conditions, this results in an enhancement in high-energy dilepton 
production; however, compared to the free-streaming case the effect is 
reduced due to the lower effective temperatures obtained by the 
collisionally-broadened plasma.  We also note that in the case of 
collisionally-broadened expansion the magnitude of $\xi$ is 
significantly reduced as compared to the free-streaming case.  As can 
be seen from the rightmost panel of Fig.~\ref{fig:modelPlotCB} even if 
one assumes a large isotropization time, $\tau_{\rm iso} = 18 \, \tau_0$, 
the amount of momentum space anisotropy generated is small with 
$\xi_{\rm max} \sim 2.5$ for $\gamma=2$ and $\xi_{\rm max} \sim 1.5$ 
for $\gamma=0.5$.


\begin{figure*}[t]
\includegraphics[width=8cm]{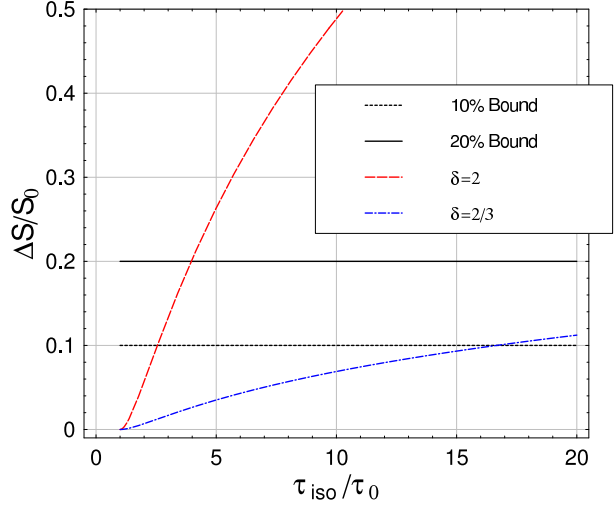}
\caption{Percentage entropy generation using fixed initial condition
interpolating models (\ref{eq:modelEQs}) with $\delta=2$ and $\delta=2/3$.
Horizontal lines show 10\% and 20\% entropy generation bounds.}
\label{fig:entropygeneration}
\end{figure*}

\begin{table}[t]
\begin{tabular}{ | l | c | c || c | c |}
\hline
{\bf Interpolating Model} & 
{\bf RHIC -- 10\%} & 
{\bf RHIC -- 20\%} & 
{\bf LHC -- 10\%} & 
{\bf LHC -- 20\%} \\ \hline
Free-Streaming ($\delta=2$) & 
$\tau_{\rm iso} \leq$ 0.8 fm/c & $\tau_{\rm iso} \leq$ 1.2 fm/c & $\tau_{\rm iso} \leq$ 0.26 fm/c & $\tau_{\rm iso} \leq$ 0.4 fm/c \\ \hline
Collisionally-Broadened ($\delta=2/3$) & 
$\tau_{\rm iso} \leq$ 5 fm/c & $\tau_{\rm iso} \leq$ 18 fm/c & $\tau_{\rm iso} \leq$ 1.6 fm/c & $\tau_{\rm iso} \leq$ 6.2 fm/c \\ \hline
\end{tabular}
\caption{Bounds on $\tau_{\rm iso}$ imposed by requiring either a 10\% or
20\% bound on percentage entropy (particle number) generation from our fixed initial condition
interpolating models.  To convert to physical
scales we have used $\tau_0 = 0.3$ fm/c for RHIC and $\tau_0 = 0.1$ fm/c for LHC.}
\label{entropygentable}
\end{table}

\subsection{Space-Time Interpolating Models with Fixed Final Multiplicity}
\label{subsec:modelmult}

In the previous subsection we constructed models which allow one to 
interpolate between an initially non-equilibrium plasma to an isotropic 
equilibrium one assuming that the initial conditions are held fixed. 
One problem with this procedure is that given fixed initial conditions 
these interpolating models will result in generation of particle 
number during the transition from $\delta\in\{2,2/3\}$ to zero.

One can derive an expression for the amount by which the number 
density is increased by starting from the general expression for the 
particle number density $n(\tau)/n_0 = \left(p_{\rm 
hard}/T_0\right)^3 (1+\xi(\tau))^{-1/2}$ and using the expression for 
$p_{\rm hard}$ derived in the previous section (\ref{pdependence}) to 
obtain
\beq
\frac{n(\tau)}{n_0} = \frac{\bar{\cal U}(\tau)}{\sqrt{1+\xi(\tau)}} \; .
\label{numberdensityfixinit}
\eeq
Taking the limit $\tau \gg \tau_{\rm iso}$ we obtain 
\beq
\lim_{\tau \gg \tau_{\rm iso}}
\frac{n(\tau)}{n_0} = \frac{\tau_{\rm iso}}{\tau} \left(\frac{\tau_{\rm iso}}{\tau_0}\right)^{\delta/2-1} 
                 \left[{\cal R}\!\left(\left(\tau_{\rm iso}/\tau_0\right)^\delta-1\right)\right]^{3/4} \; .
\label{numberdensityfixinitlimit}
\eeq
Translating this into a statement about the entropy generation using 
$S(\tau) = \tau n(\tau)$ gives
\beq
\lim_{\tau \gg \tau_{\rm iso}}
\frac{S(\tau)}{S_0} = \left(\frac{\tau_{\rm iso}}{\tau_0}\right)^{\delta/2}
 \left[{\cal R}\!\left(\left(\tau_{\rm iso}/\tau_0\right)^\delta-1\right)\right]^{3/4} \; .
\label{entropyfixinitlimit}
\eeq
When either $\delta \rightarrow 0$ or $\tau_{\rm iso} \rightarrow 
\tau_0$,  $\Delta S \equiv (S_{\rm final} - S_0)/S_0$ goes to zero and there is no entropy generation; 
however, entropy generation increases monotonically with $\delta$.  In 
the limit of large $\tau_{\rm iso}/\tau_0$ we find
\beq
\lim_{\tau_{\rm iso}\rightarrow\infty} 
\frac{\Delta S}{S_0} = \left(\frac{\tau_{\rm iso}}{\tau_0}\right)^{\frac{\delta}{8}} - 1 \; .
\label{entropylimit}
\eeq
Again we see that in the limit that either $\delta \rightarrow 
0$ or $\tau_{\rm iso} \rightarrow \tau_0$ then there is no entropy generation.

The requirement of bounded entropy generation can be used to constrain 
non-equilibrium models of the QGP \cite{Dumitru:2007qr}. In 
Fig.~\ref{fig:entropygeneration} we plot the entropy generation 
(particle number generation) resulting from our models using 
Eq.~(\ref{entropyfixinitlimit}) for $\delta \in \{2,2/3\}$ along with 
bounds at 10\% and 20\%.  In the free-streaming interpolating model 
($\delta=2$) with fixed initial conditions requiring that the 
percentage entropy generation be less than each of these bounds 
requires $\tau_{\rm iso} \leq 2.6 \, \tau_0$ for the 10\% bound and 
$\tau_{\rm iso} \leq 4 \, \tau_0$ for the 20\% bound.  In the 
collisionally-broadened interpolating model ($\delta=2/3$) with fixed 
initial conditions we obtain similarly $\tau_{\rm iso} \leq 17 \, 
\tau_0$ for the 10\% bound and $\tau_{\rm iso} \leq 62 \, \tau_0$ for 
the 20\% bound.  We summarize our results in 
Table~\ref{entropygentable}. As can be seen from 
Table~\ref{entropygentable} requiring the listed bounds on entropy 
generation the values of $\tau_{\rm iso}$ allowed in our 
free-streaming interpolating model become highly constrained. However, 
in the case of the collisionally-broadened interpolating model the 
upper-bounds imposed on $\tau_{\rm iso}$ are much larger due to the 
much lower entropy generation required to transition from 
collisionally-broadened evolution to hydrodynamic evolution.

One problem with our fixed initial condition family of models is that 
due to the fact that they generate additional particles the 
multiplicity of final particles is not independent of the assumed 
value of $\tau_{\rm iso}$. Because most of the experimental results for 
dilepton spectra are binned with respect to a fixed final multiplicity 
this means that we should also construct models which always result in 
a fixed final number density. In the following subsection we will show 
how this can be accomplished.

\begin{figure*}[t]
\includegraphics[width=17.7cm]{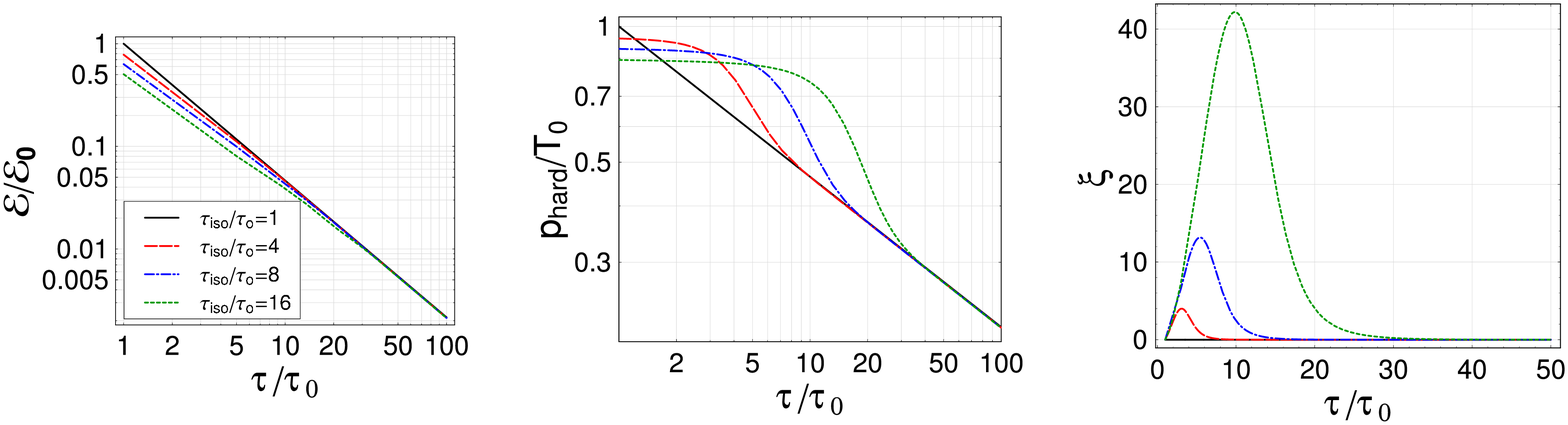}
\vspace{2mm}
\includegraphics[width=17.7cm]{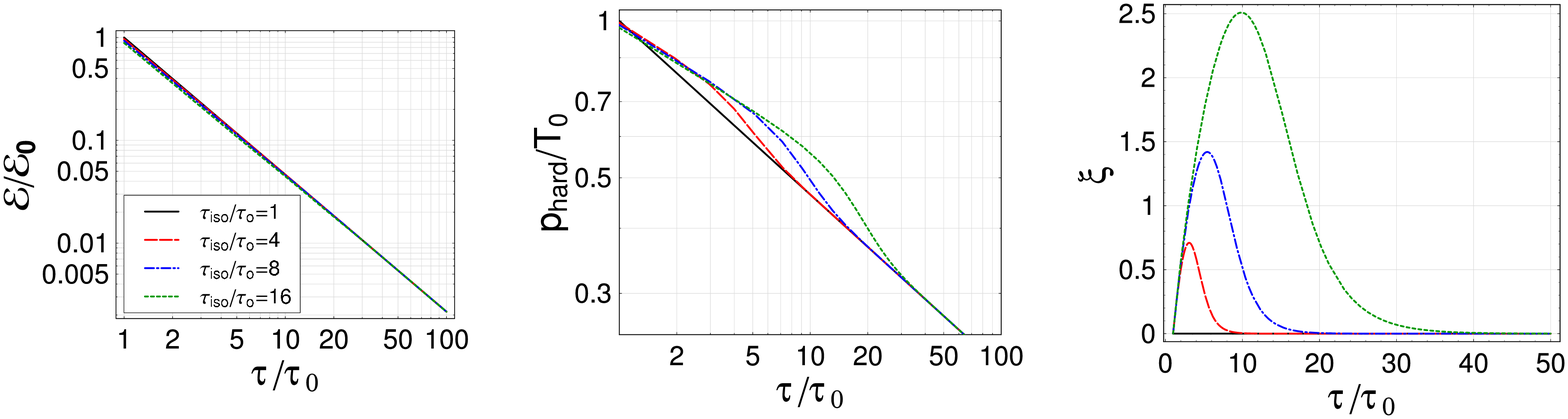}

\caption{Temporal evolution using our fixed final multiplicity interpolating models
for the energy density (left column), hard momentum 
scale (middle column), and anisotropy parameter (right column) for 
four different isotropization times $\tau_{\rm iso} \in \{1,4,6,18\} 
\, \tau_0$. Top row is the free-streaming interpolation model ($\delta=2$)
and bottom row is the collisionally-broadening interpolation model ($\delta=2/3$).
  To convert to physical scales use
$\tau_0 \sim 0.3$ fm/c for RHIC and $\tau_0 \sim 0.1$ fm/c for LHC.}

\label{fig:modelPlotFixedMult}
\end{figure*}

\subsubsection{Enforcing Fixed Final Multiplicity}

We will now construct interpolating models which have a fixed final 
entropy (multiplicity).  In order to accomplish this the initial 
conditions will have to vary as a function of the assumed 
isotropization time. We will show that, as a result, for finite 
$\tau_{\rm iso}$ one must lower the initial ``temperature'' in both 
the free-streaming and collisionally-broadened interpolating models. 
To accomplish this requires only a small modification to the 
definition of $\bar{\cal U}$ in Eq.~(\ref{Udeff}):
\begin{subequations}
\bqa
\bar{\cal U}(\tau) &\equiv& {\cal U}(\tau) \, / \, {\cal U}(\tau_{\rm iso}^+) \; , \\
{\cal U}(\tau_{\rm iso}^+) &\equiv& \lim_{\tau\rightarrow\tau_{\rm iso}^+} {\cal U}(\tau) 
  = \left[{\cal R}\!\left(\left(\tau_{\rm iso}/\tau_0\right)^\delta-1\right)\right]^{3/4} 
	  \left(\frac{\tau_{\rm iso}}{\tau_0}\right) \; .
\eqa
\label{UdeffFixMult}
\end{subequations}
As a consequence of this modification, the initial energy 
density and hence initial ``temperature'' will depend on the assumed value for 
$\tau_{\rm iso}$.  There is no modification required for $\xi$. We 
demonstrate this in Fig. \ref{fig:modelPlotFixedMult} where we plot the 
time-dependence of ${\cal E}$, $p_{\rm hard}$, and $\xi$ for 
$\gamma=2$ and (top) $\delta=2$ and (bottom) $\delta=2/3$.

In the remainder of this work we present our final results for 
dilepton yields using both approaches, i.e., fixed initial conditions 
using Eqs.~(\ref{eq:modelEQs}) with (\ref{Udeff}) or fixed final 
multiplicity through Eq.~(\ref{eq:modelEQs}) with 
(\ref{UdeffFixMult}). We mention that in both cases, dilepton 
production is affected in the presence of anisotropies in 
momentum-space, however, one anticipates that the effect will be 
larger when the initial conditions are held fixed due to the larger
particle number generation.  We will come back to this issue in 
the conclusions and discussion.


\section {Results}
\label{sec:results}

In this section we will present expected $e^+e^-$ yields resulting from 
a central Au-Au collision at RHIC full beam energy, $\sqrt{s}$=200 GeV 
and from a Pb-Pb collision at LHC full beam energy, $\sqrt{s}$=5.5 
TeV.  In all figures in this section we will present the prediction 
for RHIC energies in the left panel and LHC energies in the right 
panel.

Before presenting our results we first explain the setup, numerical 
techniques used, and parameters chosen for our calculations. Because 
the differential dilepton rate $dR^{l^+l^-}/d^4P$ given in 
Eq.~(\ref{scattering}) is independent of the assumed space-time model. 
We first evaluate it numerically using double-exponential integration 
with a target precision of $10^{-9}$.  The result for the rate was 
then tabulated on a uniformly-spaced 4-dimensional grid in $M$, $p_T$, 
$y$, and $\log_{10}\xi$ \, : $M/p_{\rm hard}, p_T/p_{\rm hard} \in 
\{0.1,25\}$, $y \in \{-3,3\}$, $\log_{10}\xi \in \{-6,4\}$.  This 
table was then used to build a four-dimensional interpolating function 
which was valid at continuous values of these four variables. We then 
boost this rate from the local reference frame to center-of-mass frame 
and evaluate the remaining integrations over space-time ($\tau$ and 
$\eta$) and transverse momentum or invariant mass appearing in 
Eqs.~(\ref{spectrumeqs}) using quasi-Monte Carlo integration with 
$\tau \in \{\tau_0,\tau_f\} $, $\eta \in \{-2.5,2.5\}$ and, depending 
on the case, restrict the integration to any cuts specified in $M$ or 
$p_T$. Our final integration time, $\tau_f$, is set by solving 
numerically for the point in time at which the temperature in our 
interpolating model is equal to the critical temperature, i.e. $p_{\rm 
hard}(\tau_f) = T_C$. We will assume that when the system reaches 
$T_C$, all medium emission stops. We are not taking into account the 
emission from the mixed/hadronic phase at late times since the 
kinematic regime we study (high $M$ and $p_T$) is dominated by 
early-time high-energy dilepton emission 
\cite{Strickland:1994rf,Mauricio:2007vz}.

For RHIC energies we take an initial temperature $T_0$= 370 MeV, at a 
formation time of $\tau_0$= 0.26 fm/c, and use $R_T$= 6.98 fm. For LHC 
energies, we use $\tau_0$= 0.088 fm/c, $T_0$= 845 MeV and $R_T$= 7.1 
fm. In both cases, the critical temperature $T_C$ is taken as 160 MeV and the spectra are calculated at midrapidity region $y=0$. 
Any cuts in transverse momentum or invariant mass will be indicated 
along with results. Note that the precise numerical value of the 
parameters above were chosen solely in order to facilitate 
straightforward comparisons with previous works \cite{Turbide:2006mc} 
from which we have obtained predictions for Drell Yan, heavy quark, 
jet-fragmentation, and jet-thermal dilepton yields. 

Finally we note that below we will use $K$-factors to adjust for 
next-to-leading order corrections to the dilepton rate.  These 
$K$-factors are determined by computing the ratio of the 
next-to-leading order prediction of \cite{Arnold:2002ja,%
Turbide:2006mc} with our leading order prediction in the case of 
ideal-hydrodynamic expansion.  We therefore assume that the 
$K$-factors are independent of the assumed thermalization time. This 
is an approximation which, in the future, one would like to relax by 
computing the full next-to-leading order dilepton rate in the presence 
of momentum-space anisotropies.

\subsection{Dilepton production with fixed initial conditions}
\label{subsec:fixinitialcond}

We now present the results of the dilepton production assuming the 
time dependence of the energy density, the hard momentum scale and the 
anisotropy parameter are given by Eqns.~(\ref{eq:modelEQs}) and 
(\ref{Udeff}) with $\delta \in \{2,2/3\}$.

\subsubsection{Free streaming interpolating model}
\label{subsubsec:fixfreestream}

\begin{figure*}[t]
\includegraphics[width=8.75cm]{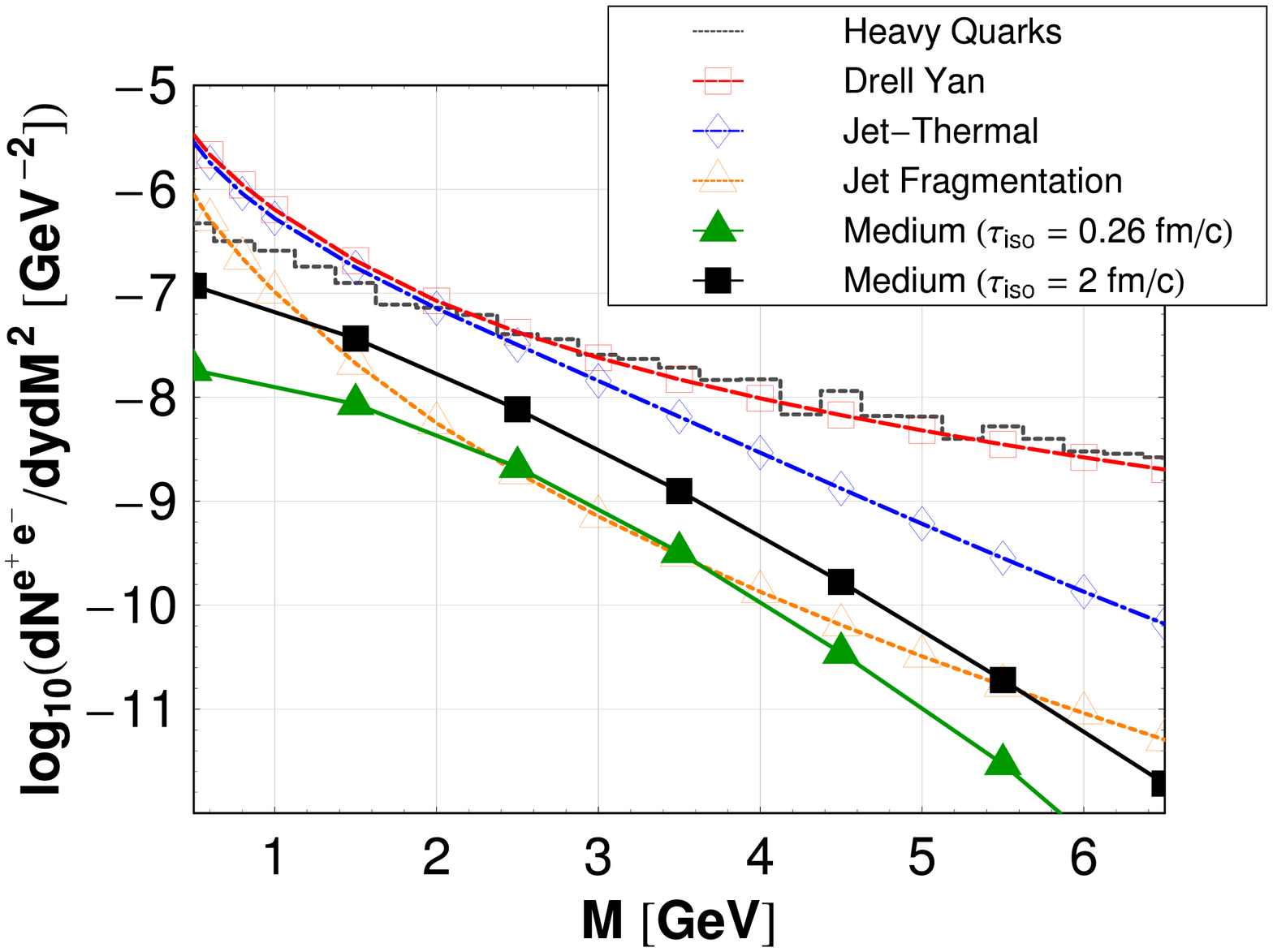}\quad
\includegraphics[width=8.75cm]{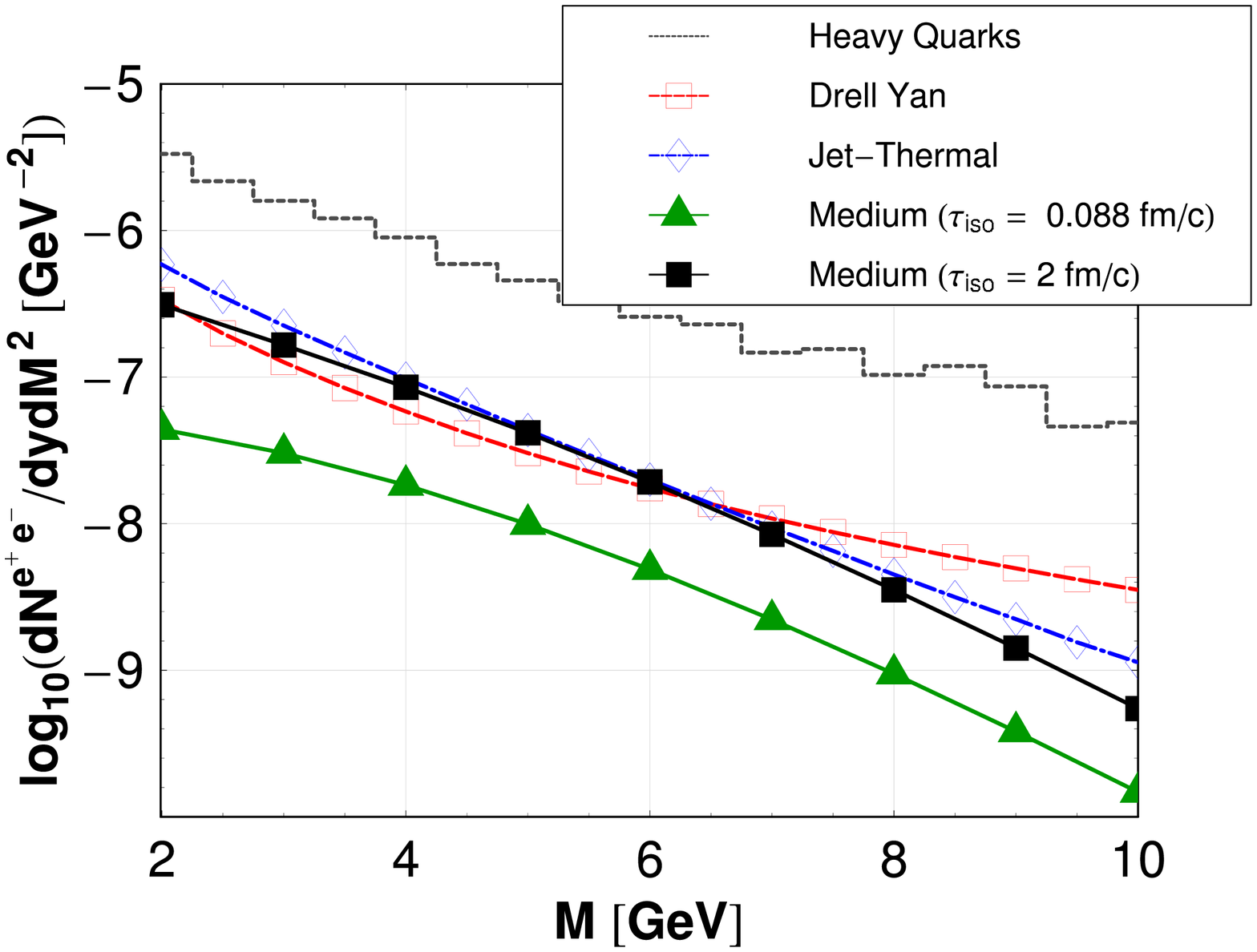}
\vspace{-2mm}
\caption{Free-streaming interpolating model dilepton yields as a 
function of invariant mass in central Au+Au collisions at RHIC (left) 
and Pb+Pb at the LHC (right), with a cut $p_T\,\geq$ 4 (8) GeV and rapidity $y$=0. For 
medium dileptons we use $\gamma$=2 and $\tau_{\rm iso}$ is taken to be 
either 0.26 (0.088) fm/c or 2 fm/c for RHIC (LHC) energies with fixed 
initial conditions. A $K$-factor of 1.5 was applied to account for NLO 
corrections. Dilepton yields from Drell Yan, Heavy Quarks, Jet-Thermal and Jet-Fragmentation were obtained from Ref.~\cite{Turbide:2006mc}.}
\label{dilmassinitial}
\end{figure*}
\begin{figure*}[t]
\includegraphics[width=8.75cm]{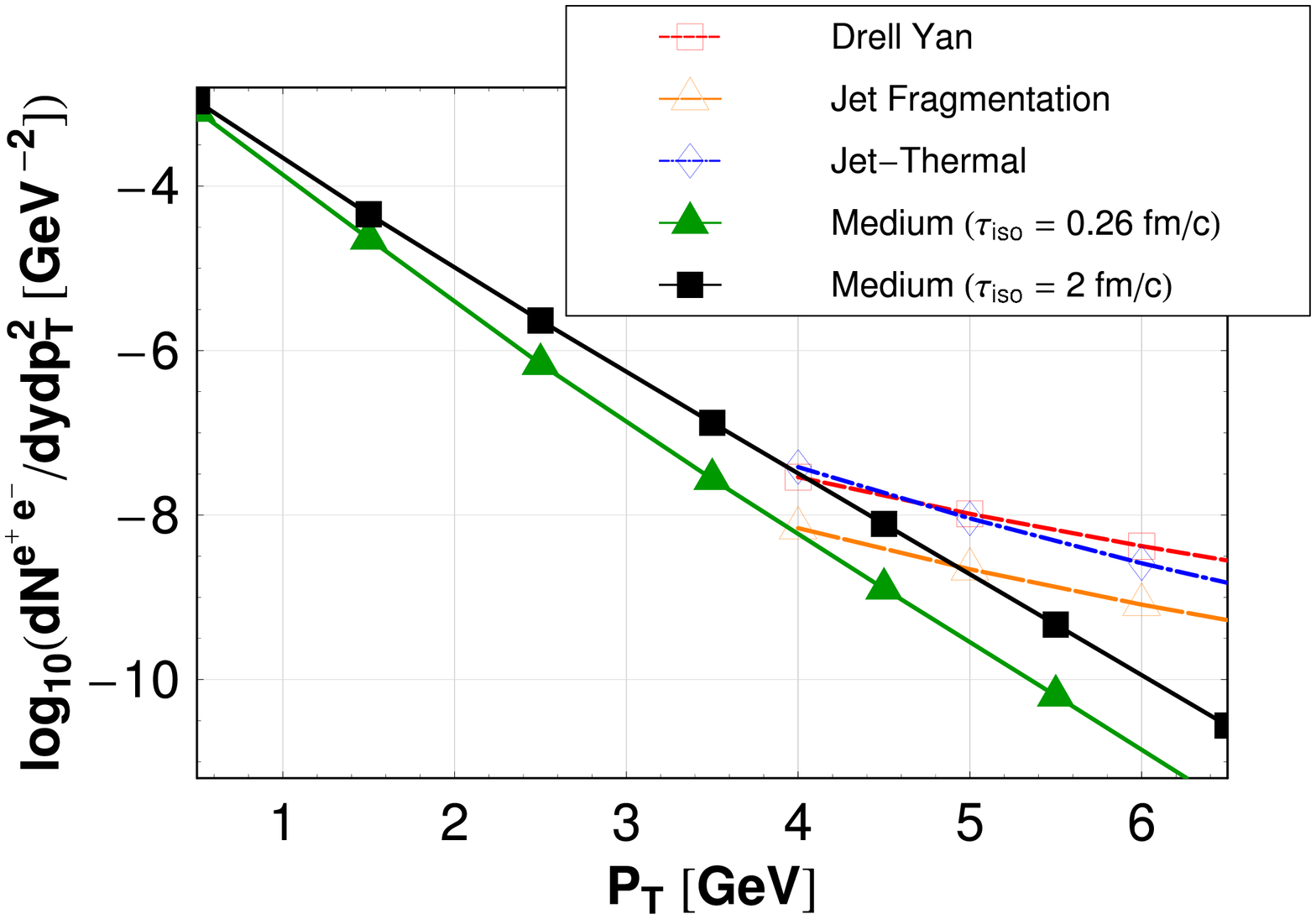}\quad
\includegraphics[width=8.75cm]{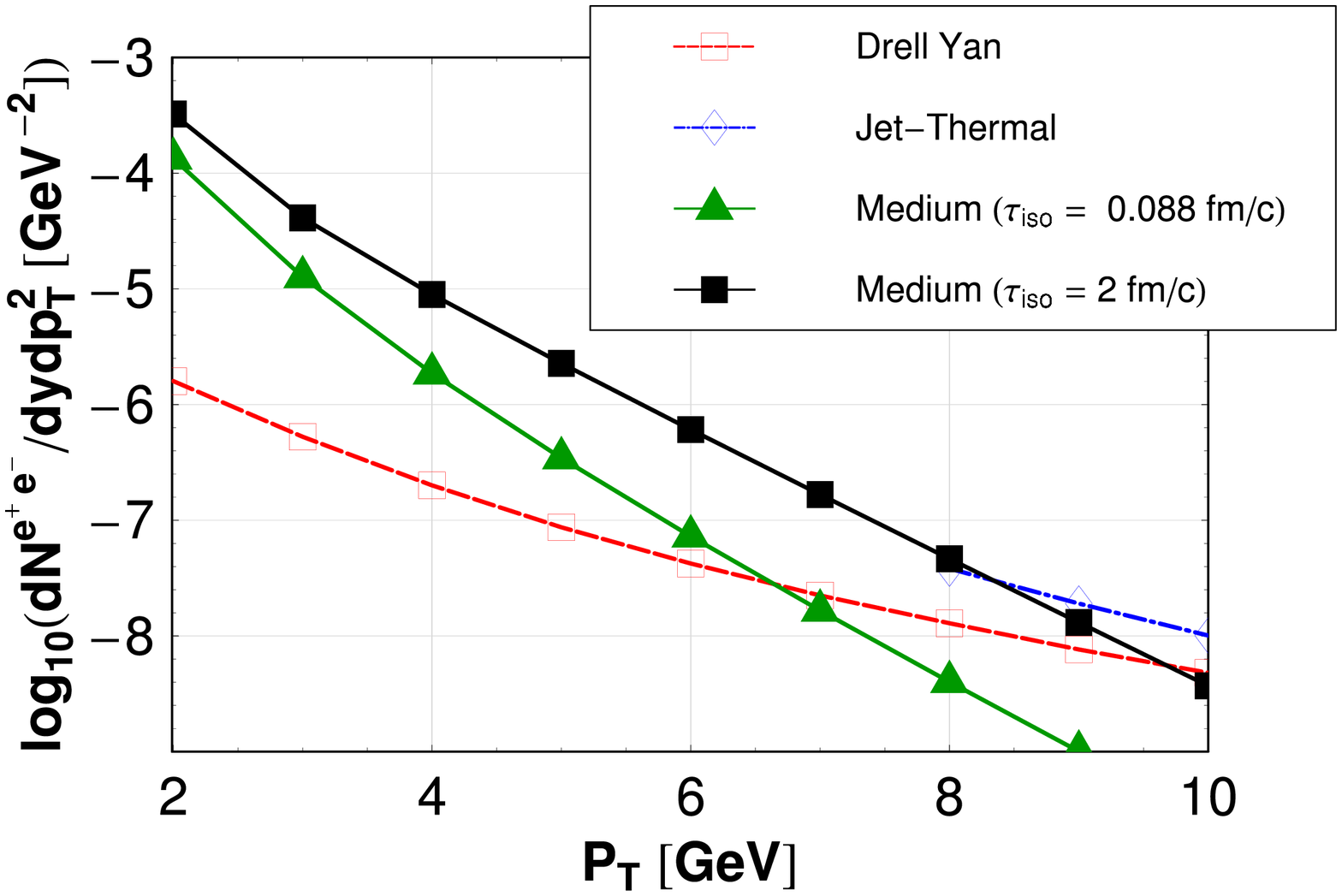}
\vspace{-2mm}
\caption{Free-streaming interpolating model dilepton yields as a 
function of transverse momentum in central Au+Au collisions at RHIC 
(left) and Pb+Pb at the LHC (right), with a cut $0.5\,\leq\,M\,\leq\,1$ GeV and rapidity $y$=0. 
For medium dileptons we use $\gamma$=2 and $\tau_{\rm iso}$ is taken 
to be either 0.26 (0.088) fm/c or 2 fm/c for RHIC (LHC) energies with 
fixed initial conditions. A $K$-factor of 6 was applied to account for 
NLO corrections. Dilepton yields from Drell Yan, Jet-Thermal and Jet-Fragmentation were obtained from Ref.~\cite{Turbide:2006mc}.}
\label{dilptinitial}
\end{figure*}

\begin{figure*}
\includegraphics[width=8.75cm]{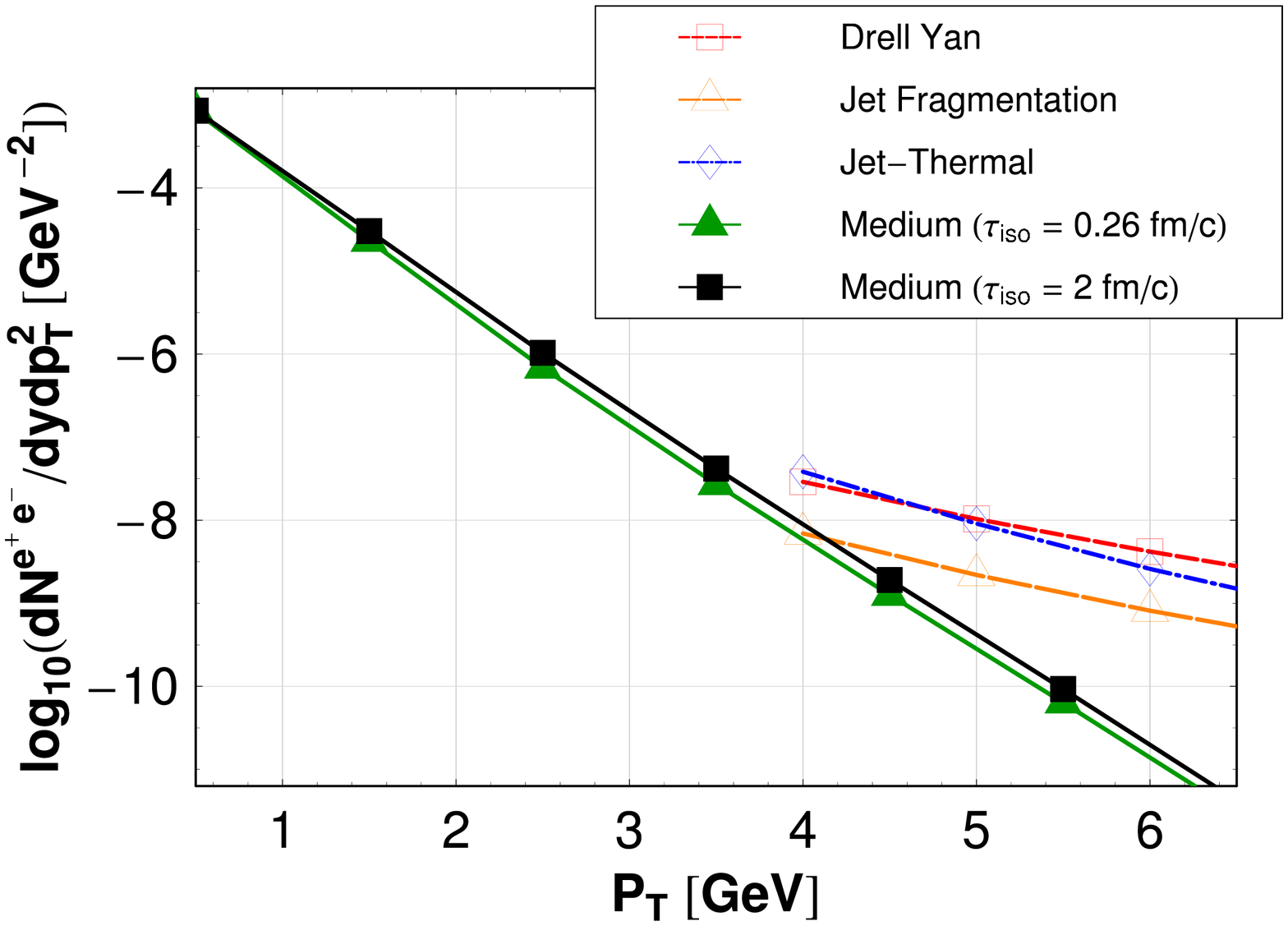}\quad
\includegraphics[width=8.75cm]{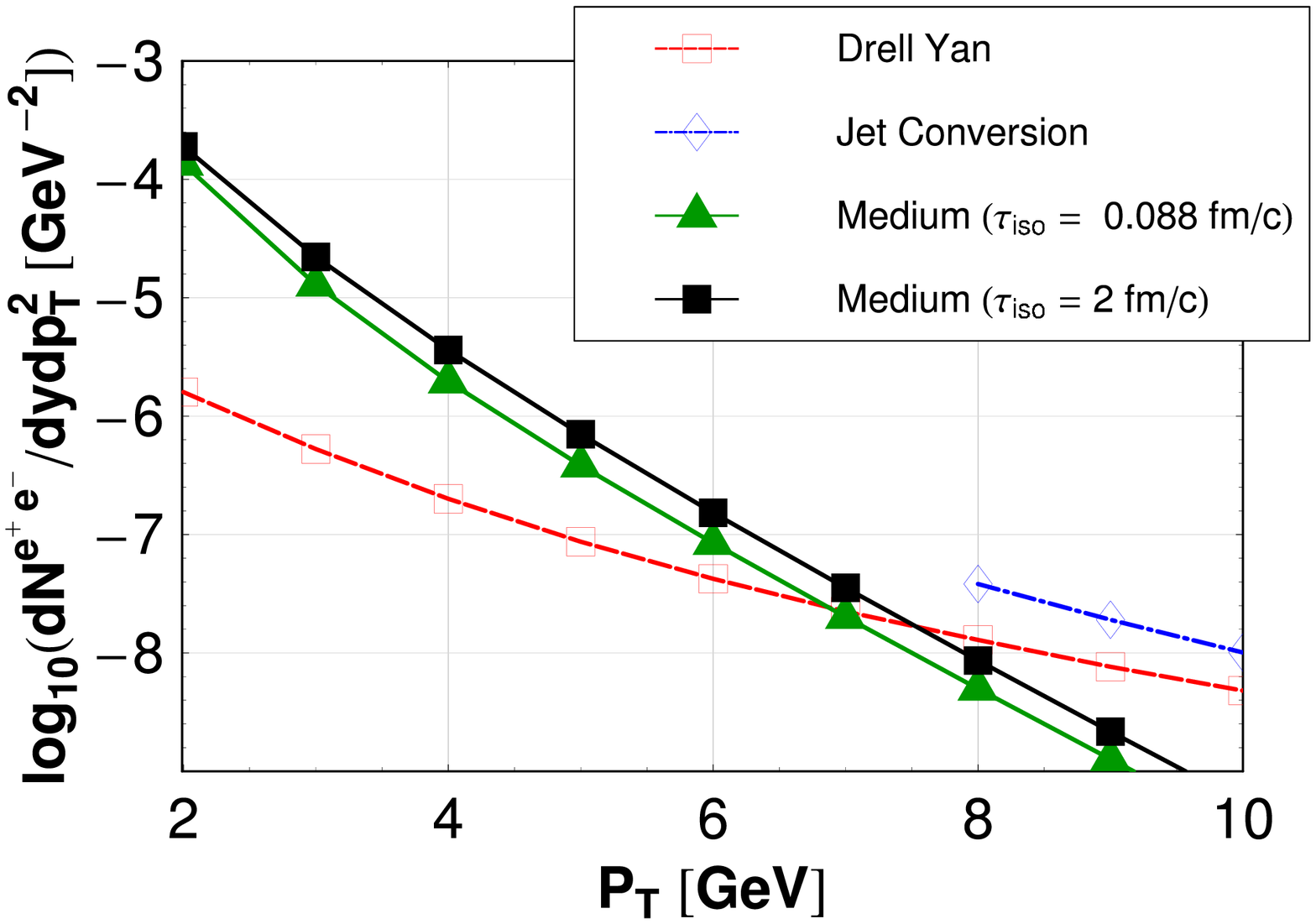}
\vspace{-2mm}
\caption{Collisionally-broadened interpolating model dilepton yields including collisional broadening as a 
function of transverse momentum in central Au+Au collisions at RHIC 
(left) and Pb+Pb at the LHC (right), with a cut $0.5\,\leq\,M\,\leq\,1$ GeV and rapidity $y$=0. 
For medium dileptons we use $\gamma$=2 and $\tau_{\rm iso}$ is taken 
to be either 0.26 (0.088) fm/c or 2 fm/c for RHIC (LHC) energies with fixed initial conditions. A 
$K$-factor of 6 was applied to account for NLO corrections. Dilepton yields from Drell Yan, Jet-Thermal and Jet-Fragmentation were obtained from Ref.~\cite{Turbide:2006mc}.}
\label{dilptptbroad}
\end{figure*}
\begin{figure*}[t]
\includegraphics[width=8.5cm]{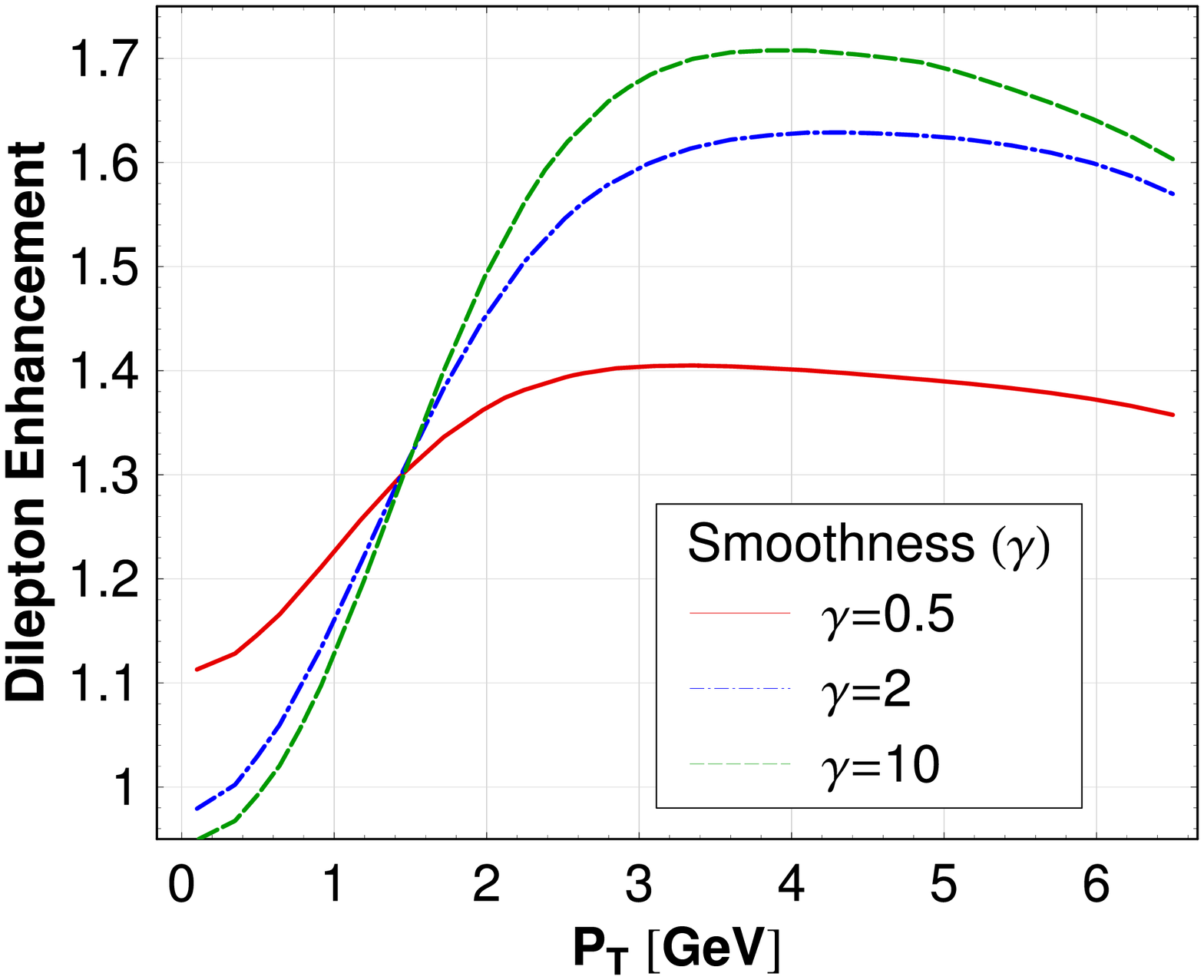}\quad
\includegraphics[width=8.5cm]{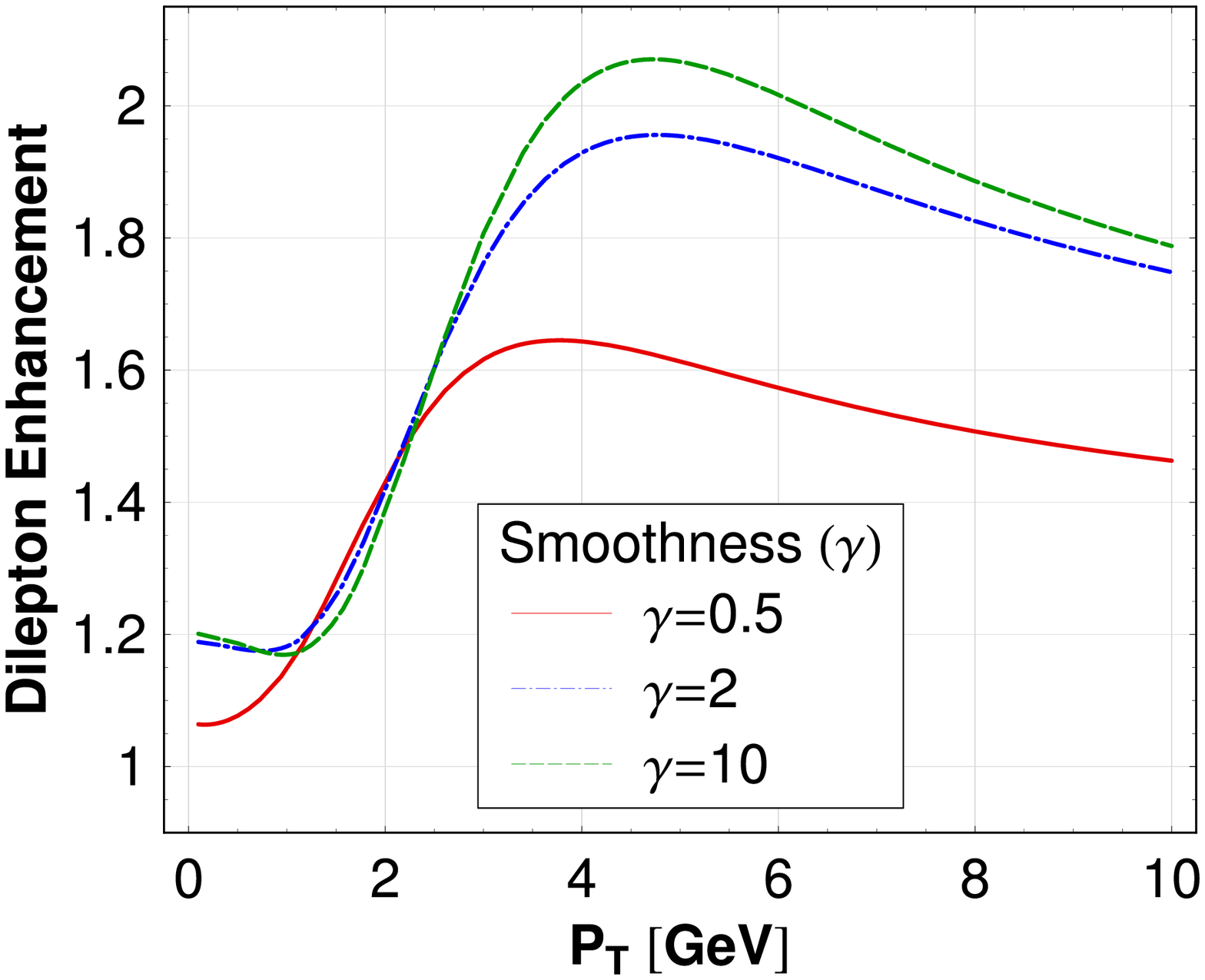}
\vspace{-2mm}
\caption{ 
Dilepton enhancement, $\phi$, as defined in Eq.~(\ref{dileptonenhancement}) 
resulting from our collisionally-broadened interpolating model 
($\delta=2/3$) with fixed initial conditions and $\tau_{\rm iso}=$ 2 
fm/c.  Result for RHIC energies is shown on left and for LHC energies 
on right.  The invariant mass cut used was $0.5\,\leq\,M\,\leq\,1$ GeV and rapidity $y$=0.
Lines show expected pre-equilibrium dilepton enhancements 
for different values of the transition width $\gamma$ corresponding to 
sharp or smooth transitions between pre-equilibrium and equilibrium 
behavior.
}
\label{enhance}
\end{figure*}

In Fig. \ref{dilmassinitial}, we show our predicted dilepton mass 
spectrum for RHIC and LHC energies assuming the time dependence of the 
energy density, the hard momentum scale and the anisotropy parameter 
are given by Eqns.~(\ref{eq:modelEQs}) and (\ref{Udeff}) with 
$\delta=2$.  This corresponds to our free-streaming interpolating 
model. For us this model will serve as an upper-bound on the possible 
effect of momentum-space anisotropies on dilepton yields. From Fig. 
\ref{dilmassinitial} we see that for both RHIC or LHC energies, there 
is a significant enhancement of up to one order of magnitude in the 
medium dilepton yield when we vary the isotropization time from 
$\tau_0$ to 2 fm/c. This enhancement is due to the fact that in 1+1 
dimensional free streaming, the system preserves more transverse 
momentum as can be seen from Fig.~\ref{fig:modelPlotFS}.  For fixed 
initial conditions this results in a larger effective temperature than 
would be obtained if the system underwent locally-isotropic 
(hydrodynamical) expansion throughout its evolution.

Nevertheless, as Fig. \ref{dilmassinitial} shows, as a function of 
invariant mass, the other contributions to high-energy dilepton yields 
(Drell-Yan, jet-thermal, and jet-fragmentation) are all of the same 
order of magnitude as the medium contribution.  This coupled with the 
large background coming from semileptonic heavy quarks decays would 
make it extremely difficult for experimentalists to extract a clean 
medium dilepton signal from the invariant mass spectrum.  For this 
reason it does not look very promising to determine plasma initial 
conditions from the dilepton invariant mass spectrum.  For this reason 
we will not present our predictions for the invariant mass spectrum 
for the intermediate models detailed below and only return to the 
invariant mass spectrum at the end of this section for completeness.

The good news is, however, that as a function of transverse momentum, 
see Fig. \ref{dilptinitial}, the production of medium dileptons is 
expected to dominate other production mechanisms for $p_T \lsim$ 4 (6) 
GeV in the case of RHIC (LHC). In addition to this we see that for the 
free-streaming interpolating model that there is a significant 
enhancement of medium dileptons for both RHIC and LHC energies.

In order to quantify the effect of time-dependent pre-equilibrium 
emissions we define the ``dilepton enhancement'', $\phi(\tau_{\rm 
iso})$, as the ratio of the dilepton yield obtained with an isotropization time 
of $\tau_{\rm iso}$ to that obtained from an instantaneously 
thermalized plasma undergoing only 1+1 hydrodynamical expansion, ie. 
$\tau_{\rm iso}=\tau_0$.
\beq
\phi(\tau_{\rm iso}) \equiv \left. \left( \dfrac{dN^{e^+e^-}(\tau_{\rm iso})}{dy dp_T^2} \right) \right/
\left( \dfrac{dN^{e^+e^-}(\tau_{\rm iso}=\tau_0)}{dy dp_T^2} \right)
\label{dileptonenhancement}
\eeq
Using this criterion we find for the free streaming interpolating 
model with fixed initial conditions the dilepton enhancement at 
$\tau_{\rm iso} = 2$ fm/c can be as large as 10. However, as mentioned 
above we expect that the actual enhancement will be lower due to the 
fact that parton interactions such as collisional-broadening will 
modify the free-streaming $\xi = \tau^2/\tau_0^2-1$ to something 
growing slower in proper time bringing the system closer to 
equilibrized expansion.  In addition, as we will discuss below when 
using fixed initial conditions and $\delta=2$ there is significant 
entropy generation which, when properly normalized to fixed final 
multiplicity, results in reduced $\phi$. Therefore, we expect $\phi 
\sim 10$ obtained from the free streaming interpolation model with 
fixed initial conditions to be an upper-bound on the effect of 
pre-equilibrium emissions. Some of our results fixing initial conditions 
are related with recent work on dilepton production from a viscous QGP~\cite{Dusling:2008xj}.

\subsubsection{Collisionally-broadened interpolating model}
\label{subsubsec:fixcollisional}

In Fig. \ref{dilptptbroad}, we show our predicted dilepton transverse 
momentum spectrum for RHIC and LHC energies assuming the time 
dependence of the energy density, the hard momentum scale and the 
anisotropy parameter are given by Eqns.~(\ref{eq:modelEQs}) and 
(\ref{Udeff}) with $\delta=2/3$.  This corresponds to our 
collisionally-broadened interpolating model with fixed initial 
conditions. From Fig. \ref{dilptptbroad} we see that for both RHIC or 
LHC energies, there is dilepton enhancement in the kinematic range 
shown; however, compared to the free streaming case the enhancement is 
reduced.  This is due to the fact that the collisionally-broadened 
interpolating model is always closer to locally-isotropic expansion 
than the free-streaming ($\delta=2$) model, see 
Figs.~\ref{fig:modelPlotFS} and \ref{fig:modelPlotCB}.

In Fig.~\ref{enhance} we show the dilepton enhancement, $\phi$, as 
function of transverse momentum for $\tau_{\rm iso} = 2$ fm/c at 
(left) RHIC energies (right) LHC energies.  The invariant mass cut is 
the same as in Fig.~\ref{dilptptbroad} ($0.5\,\leq\,M\,\leq\,1$ GeV).  As can 
be seen from Fig.~\ref{enhance} using fixed initial conditions there 
is a rapid increase in $\phi$ between 1 and 3 GeV at RHIC energies and 
1 and 4 GeV at LHC energies.  The precise value of the enhancement 
depends on the assumed width $\gamma^{-1}$ and in Fig.~\ref{enhance} 
we show $\phi$ for $\gamma^{-1} \in \{0.1,0.5,2\}$.  As can be seen 
from this figure both sharp and smooth transitions from early-time 
collisionally-broadened expansion to ideal hydrodynamic expansion 
result in a 40-70\% enhancement of medium dilepton yields at RHIC 
energies and 60-100\% at LHC energies.  We will return to this in the 
results summary at the end of this section.

\subsection{Dilepton production with fixed final multiplicity}
\label{subsec:fixfinalmult}

We now present the results of the dilepton production assuming the 
time dependence of the energy density, the hard momentum scale and the 
anisotropy parameter are given by Eqns.~(\ref{eq:modelEQs}) and 
(\ref{UdeffFixMult}) with $\delta \in \{2,2/3\}$.

\subsubsection{Free streaming interpolating model}
\label{subsubsec:multfreestream}

\begin{figure*}[t]
\includegraphics[width=8.65cm]{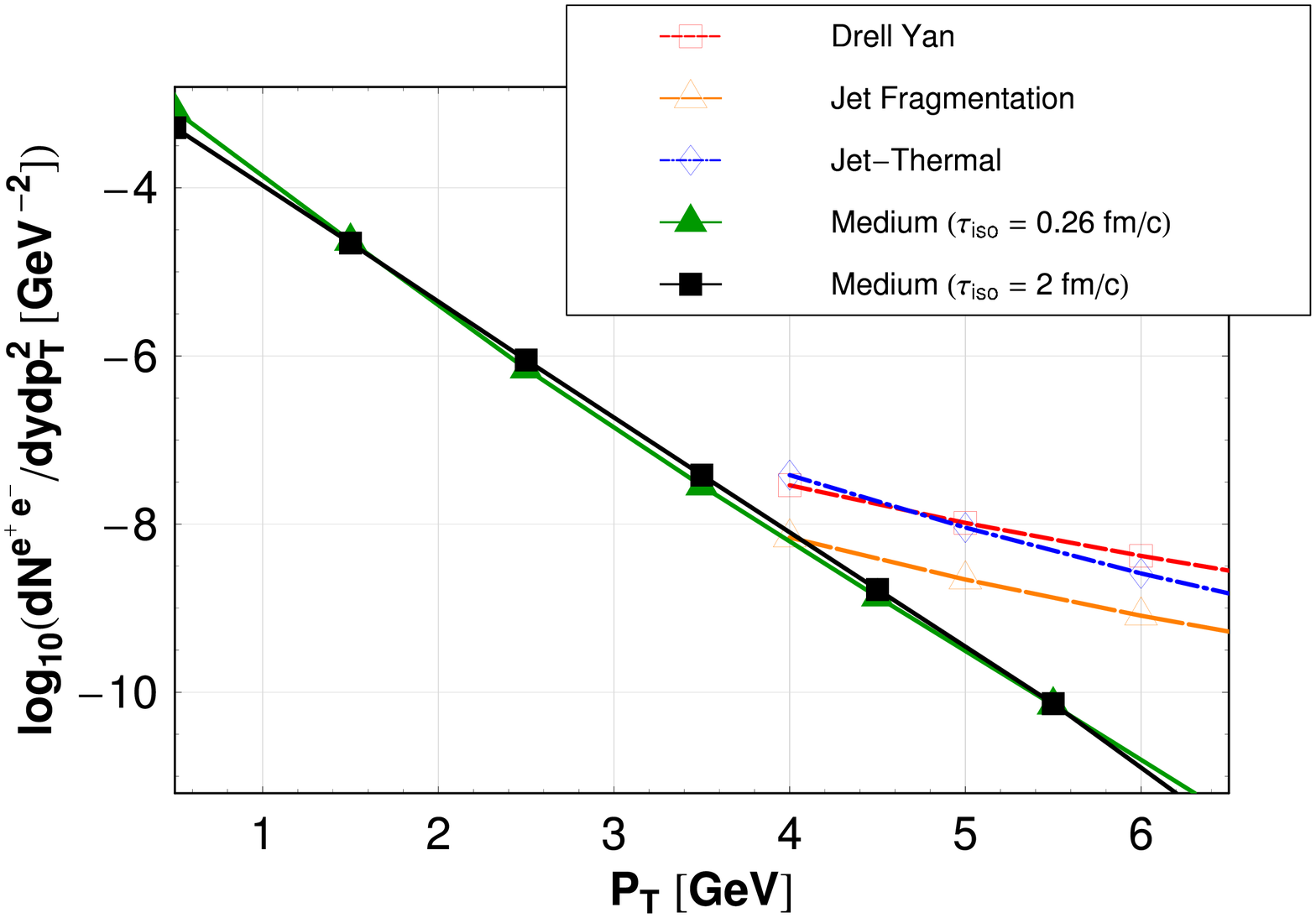}\quad
\includegraphics[width=8.75cm]{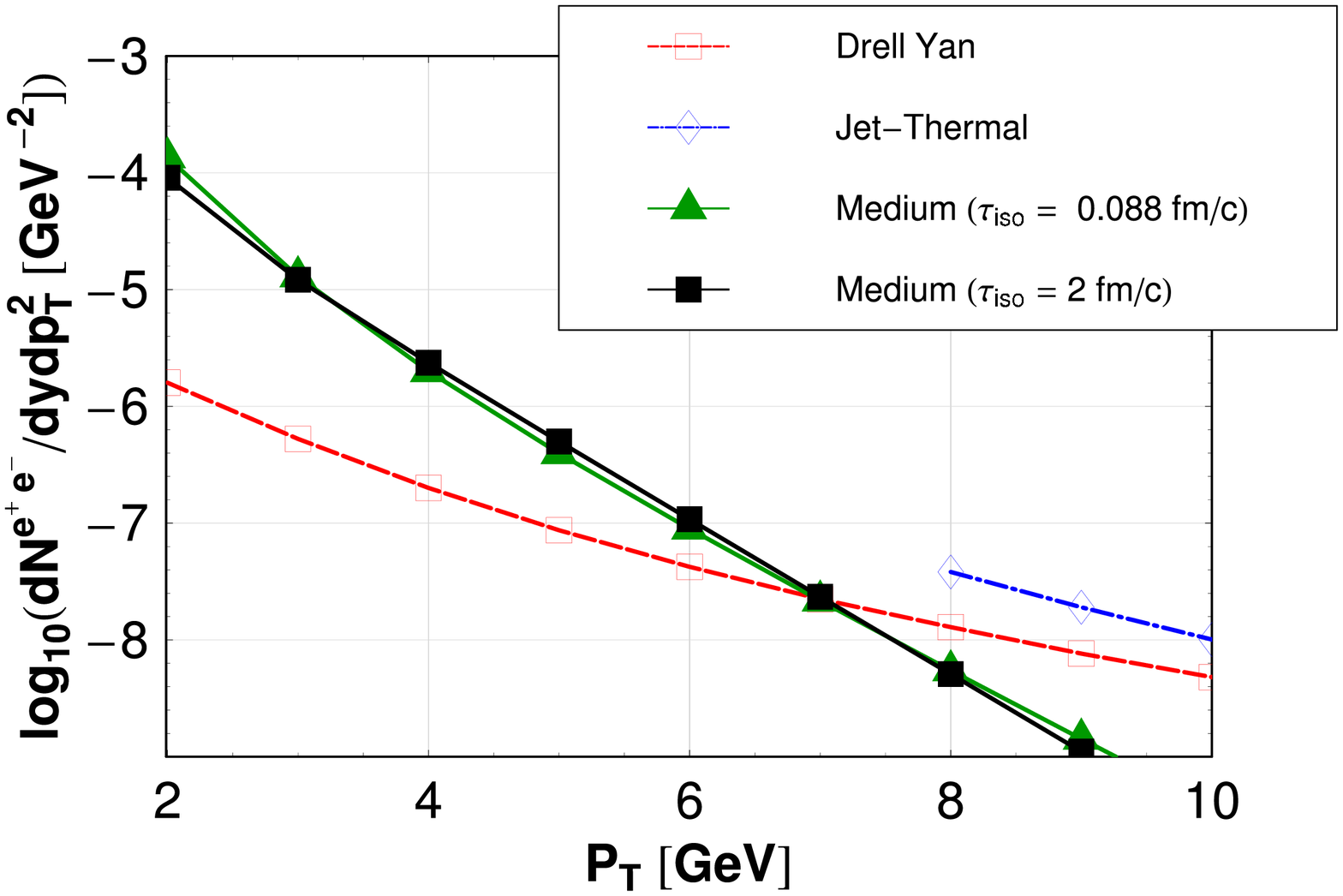}
\vspace{-2mm}
\caption{Free streaming interpolating model dilepton yields as a function of transverse momentum in 
central Au+Au collisions at RHIC (left) and Pb+Pb at the LHC (right), 
with a cut $0.5\,\leq\,M\,\leq\,1$ GeV and rapidity $y$=0. For medium dileptons we use 
$\gamma$=2 and $\tau_{\rm iso}$ is taken to be either 0.26 (0.088) 
fm/c or 2 fm/c for RHIC (LHC) energies and fixed final multiplicity. A $K$-factor of 6 was applied to 
account for NLO corrections and rapidity $y$=0. Dilepton yields from Drell Yan, Jet-Thermal and Jet-Fragmentation were obtained from Ref.~\cite{Turbide:2006mc}.}
\label{dilptmult}
\end{figure*}

In Fig. \ref{dilptmult}, we show our predicted dilepton transverse 
momentum spectrum for RHIC and LHC energies assuming the time 
dependence of the energy density, the hard momentum scale and the 
anisotropy parameter are given by Eqns.~(\ref{eq:modelEQs}) and 
(\ref{UdeffFixMult}) with $\delta=2$.  This corresponds to our 
free-streaming interpolating model with fixed final multiplicity. From 
Fig. \ref{dilptmult} we see that for both RHIC or LHC energies, there 
is dilepton enhancement in the kinematic range shown; however, when 
fixing on final multiplicity the effect of a free-streaming 
pre-equilibrium phase is reduced.  In fact, for small and large $p_T$ 
the free-streaming interpolating model with fixed final multiplicities 
predicts a suppression of dileptons.  This is due to the fact that in 
order to maintain fixed final multiplicity for $\tau_{\rm iso}=2$ fm/c 
the free-streaming model initial energy density has to be reduced by 
$\sim 50\%$ (see top row of Fig.~\ref{fig:modelPlotFixedMult}).

\subsubsection{Collisionally-broadened interpolating model}
\label{subsubsec:multcollisional}

\begin{figure*}[t]
\includegraphics[width=8.75cm]{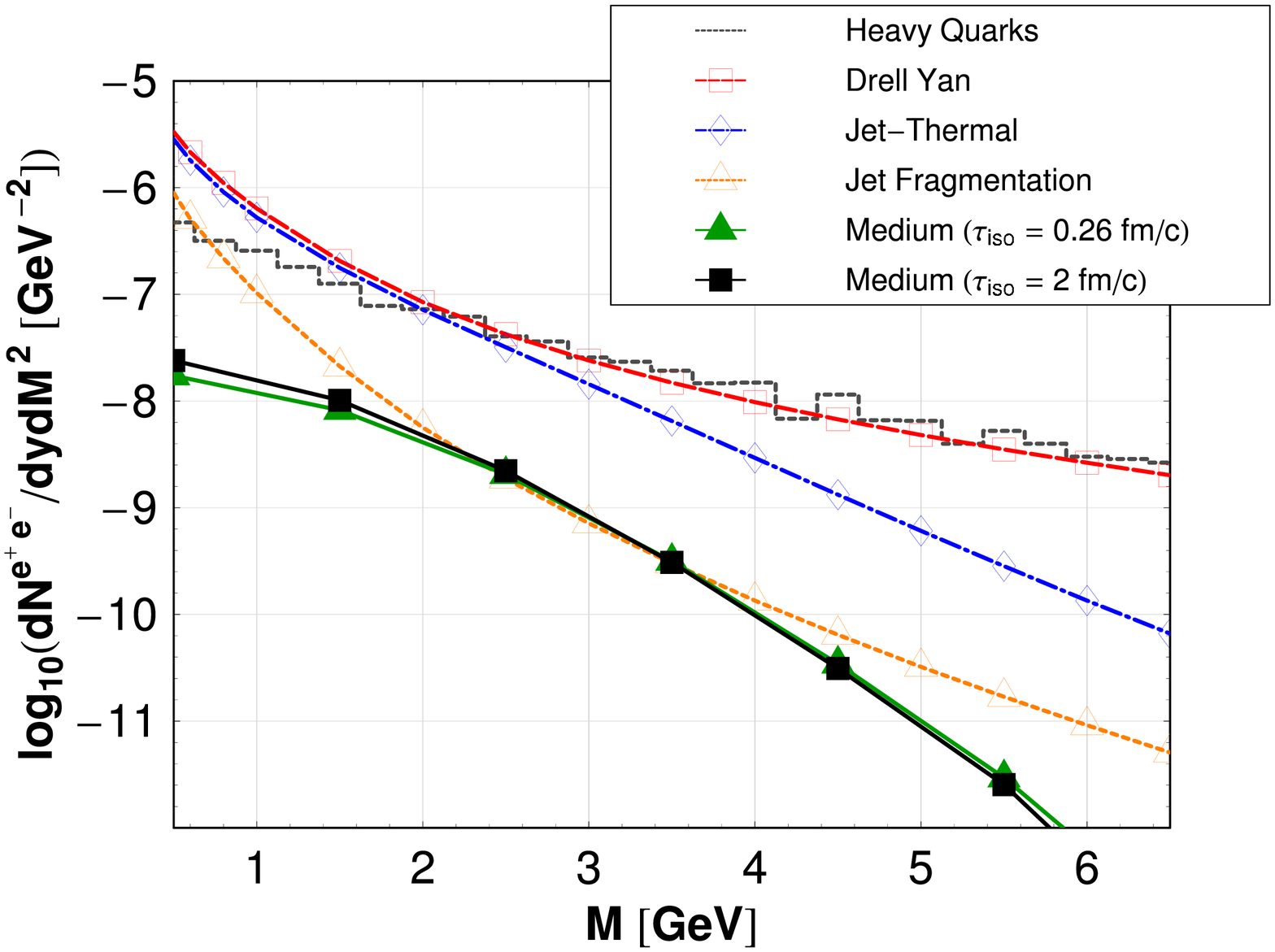}\quad
\includegraphics[width=8.75cm]{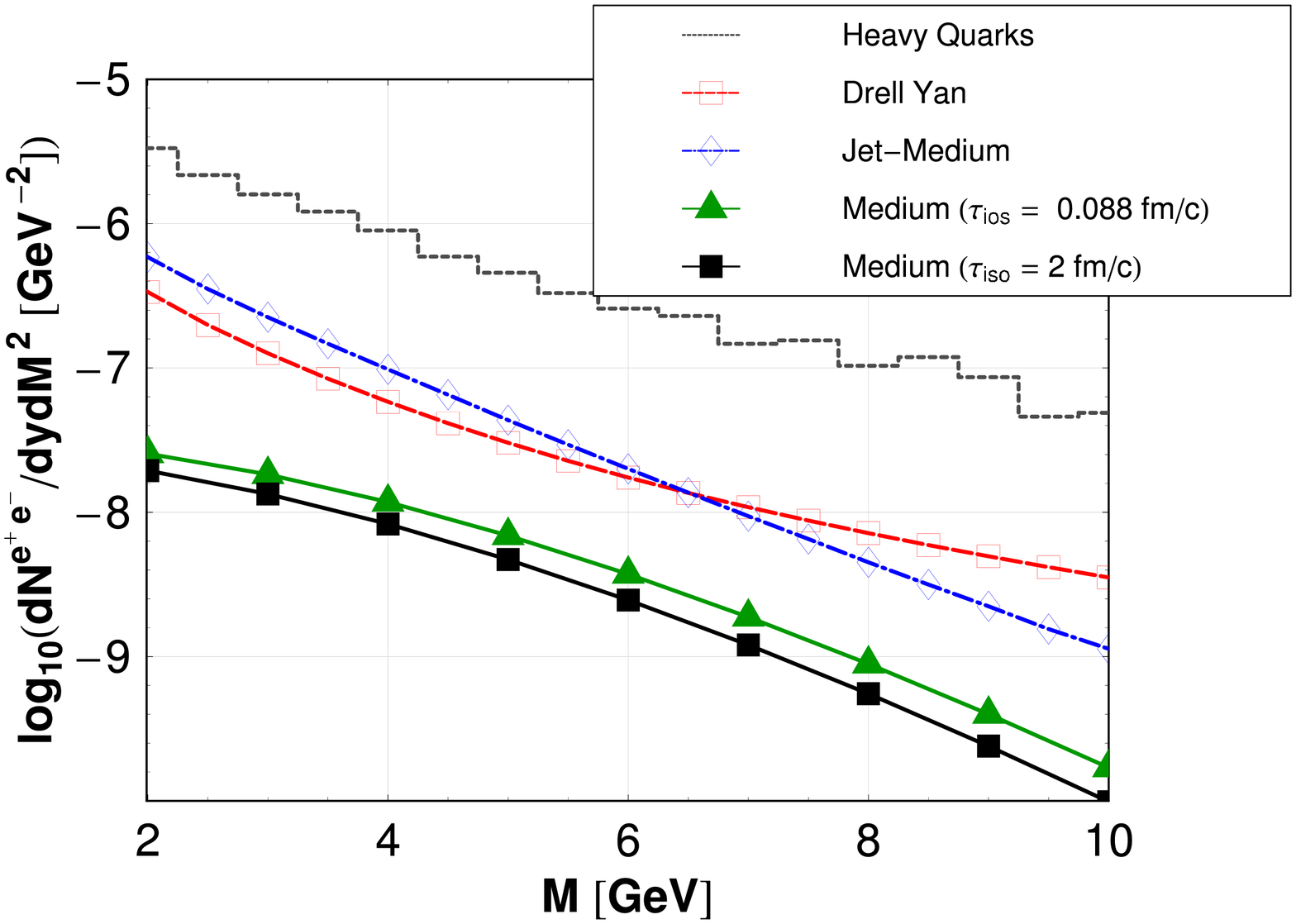}
\vspace{-2mm}
\caption{Collisionally-broadened interpolating model dilepton yields as a function of invariant mass in central 
Au+Au collisions at RHIC (left) and Pb+Pb at the LHC (right), with a 
cut $p_T\,\geq$ 4 (8) GeV and and rapidity $y$=0. For medium dileptons we use $\gamma$=2 and 
$\tau_{\rm iso}$ is taken to be either 0.26 (0.088) fm/c or 2 fm/c for 
RHIC (LHC) energies and fixed final multiplicity. A $K$-factor of 1.5 was applied to account for 
NLO corrections. Dilepton yields from Drell Yan, Heavy Quarks, Jet-Thermal and Jet-Fragmentation were obtained from Ref.~\cite{Turbide:2006mc}.}
\label{dilmassmultbroad}
\end{figure*}
\begin{figure*}[t]
\includegraphics[width=8.75cm]{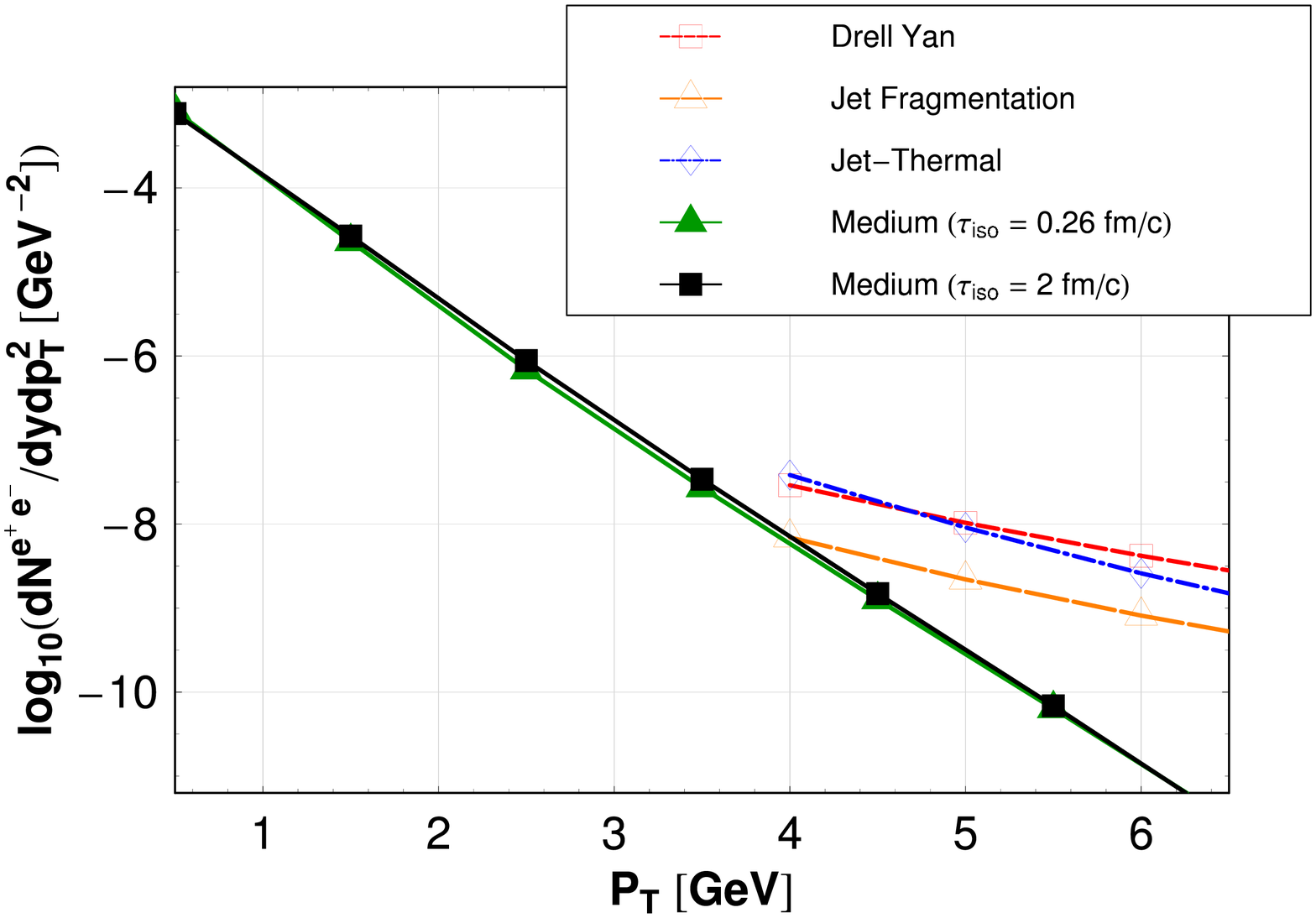}\quad
\includegraphics[width=8.75cm]{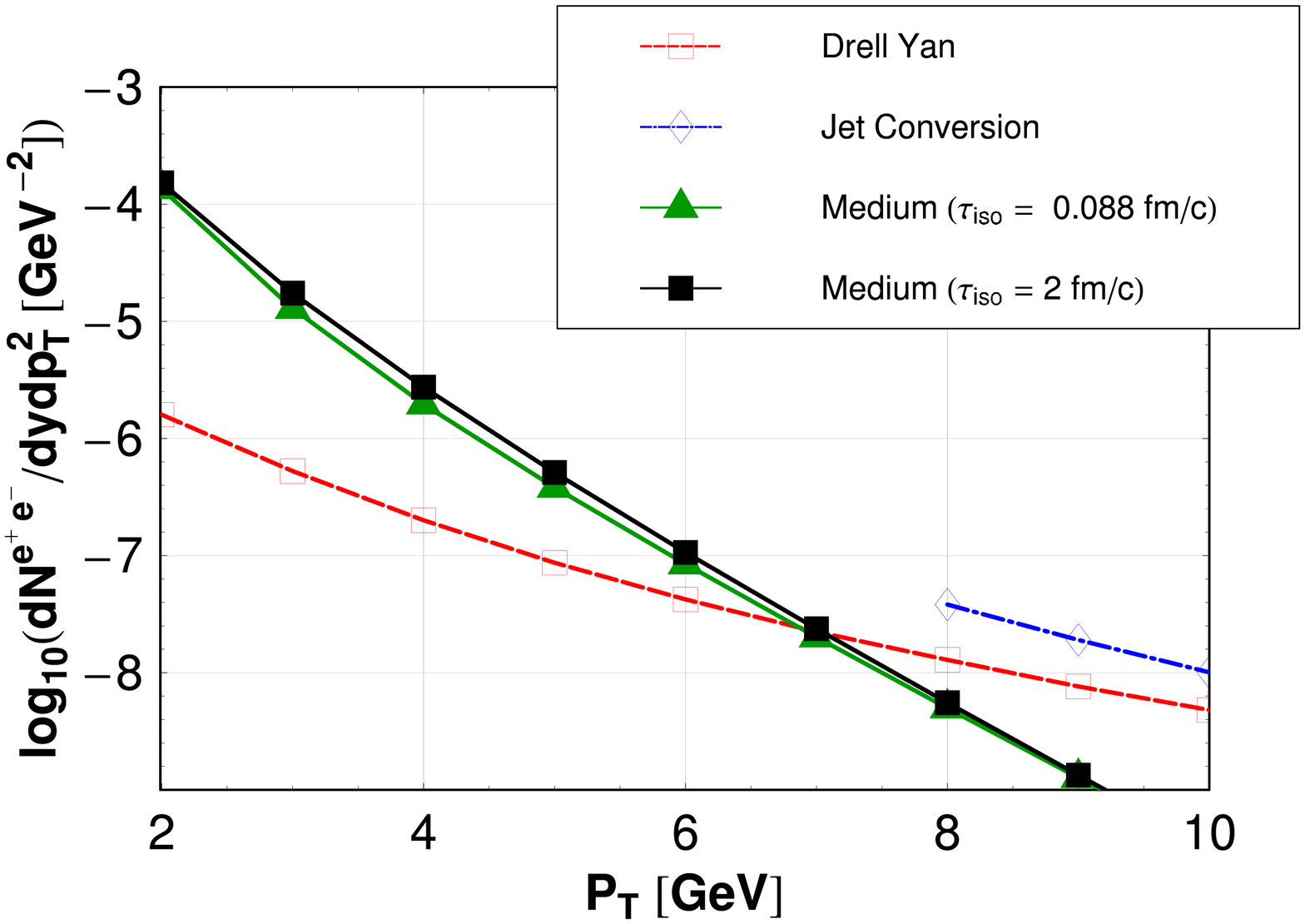}
\vspace{-2mm}
\caption{Collisionally-broadened interpolating model dilepton yields as a function of transverse momentum in 
central Au+Au collisions at RHIC (left) and Pb+Pb at the LHC (right), 
with a cut $0.5\,\leq\,M\,\leq\,1$ GeV and rapidity $y$=0. For medium dileptons we use 
$\gamma$=2 and $\tau_{\rm iso}$ is taken to be either 0.26 (0.088) 
fm/c or 2 fm/c for RHIC (LHC) energies and fixed final multiplicity. A $K$-factor of 6 was applied to 
account for NLO corrections. Dilepton yields from Drell Yan, Jet-Thermal and Jet-Fragmentation were obtained from Ref.~\cite{Turbide:2006mc}.}
\label{dilptmultbroad}
\end{figure*}
\begin{figure*}[t]
\includegraphics[width=8.5cm]{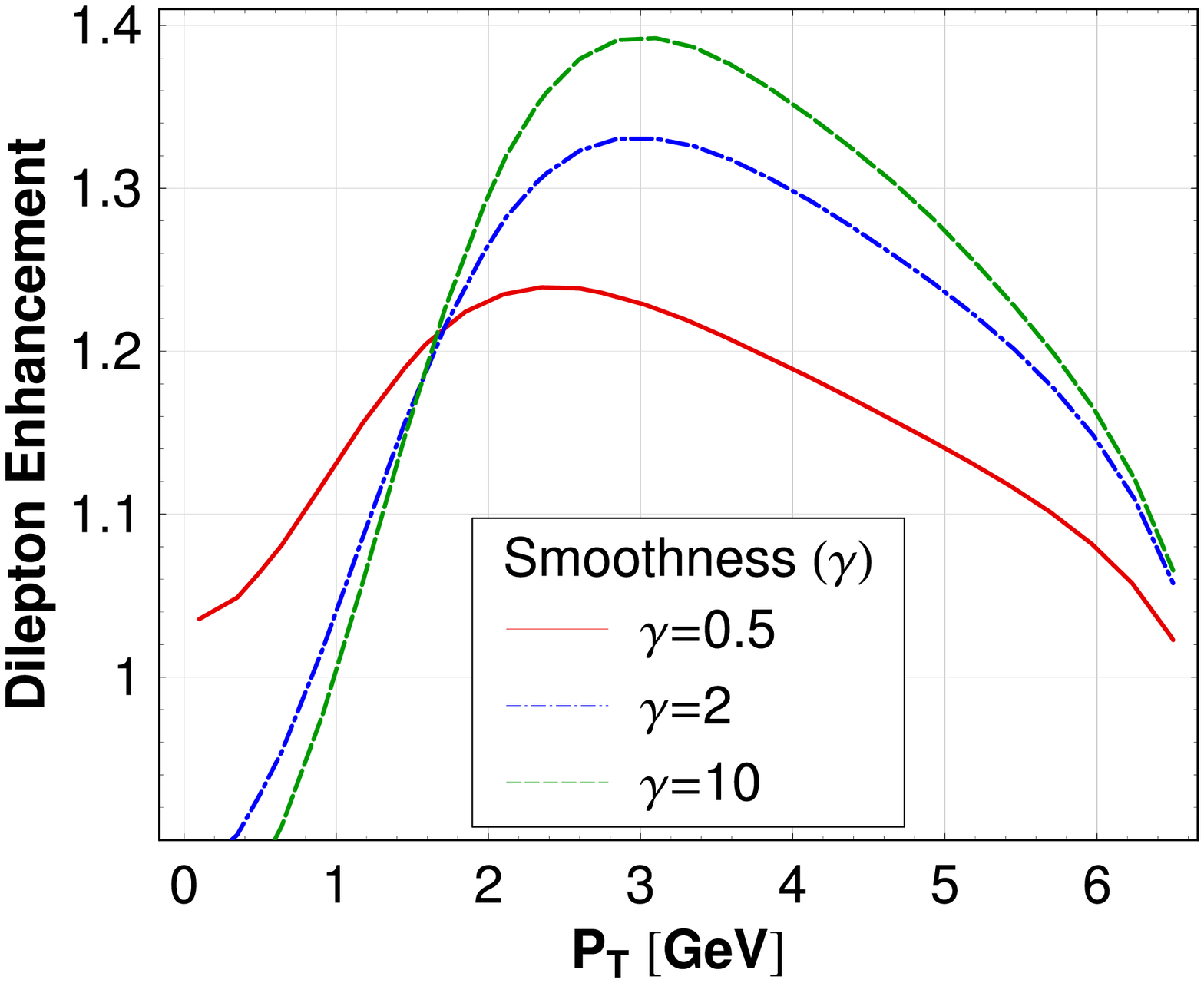}\quad
\includegraphics[width=8.5cm]{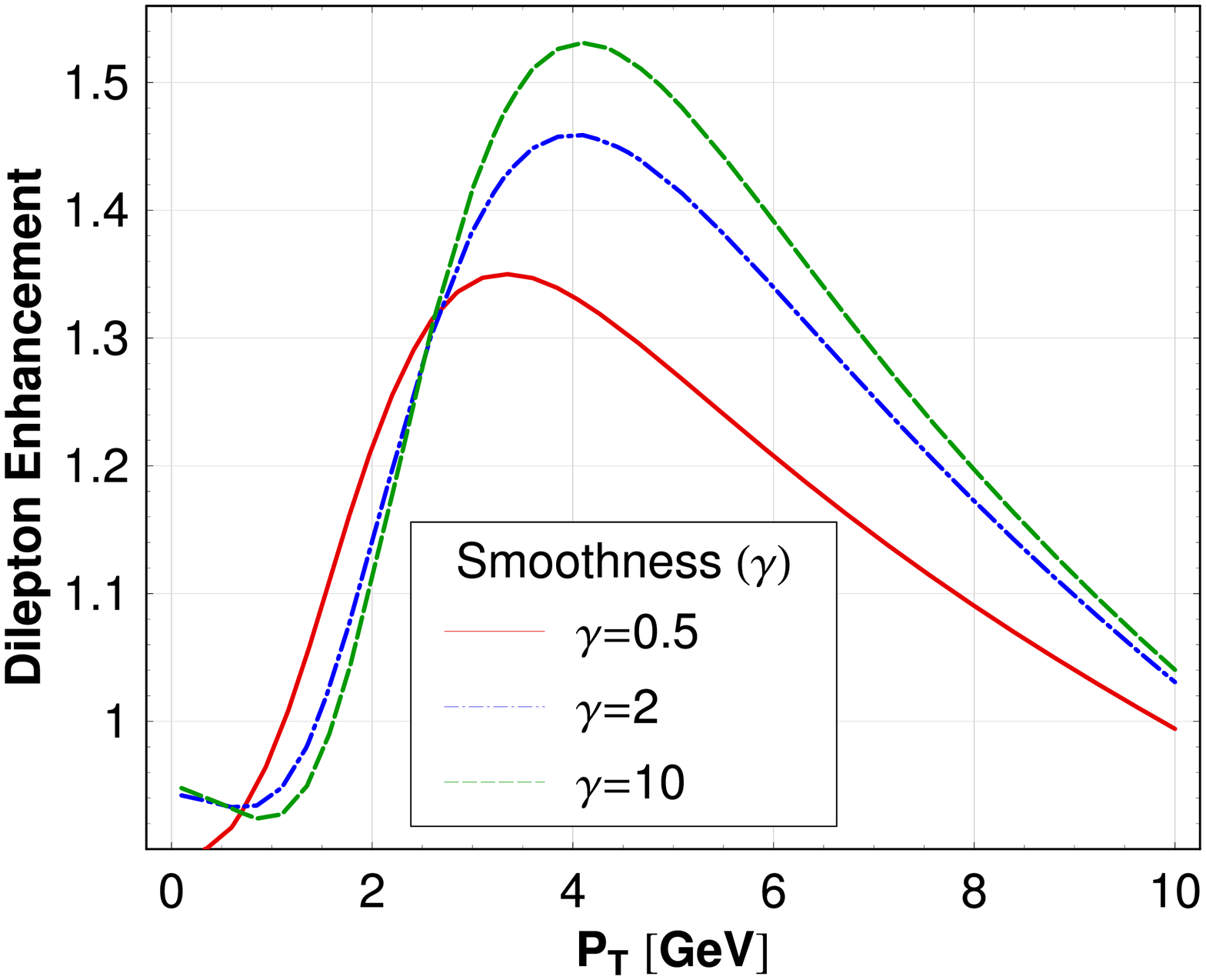}
\vspace{-2mm}
\caption{
Dilepton enhancement, $\phi$, as defined in Eq.~(\ref{dileptonenhancement}) 
resulting from our collisionally-broadened interpolating model 
($\delta=2/3$) with fixed final multiplicity and $\tau_{\rm iso}=$ 2 
fm/c.  Result for RHIC energies is shown on left and for LHC energies 
on right.  The invariant mass cut used was $0.5\,\leq\,M\,\leq\,1$ GeV and and rapidity $y$=0.
Lines show expected pre-equilibrium dilepton enhancements 
for different values of the transition width $\gamma$ corresponding to 
sharp or smooth transitions between pre-equilibrium and equilibrium 
behavior.
}
\label{enhance1-mult}
\end{figure*}

In Figs. \ref{dilmassmultbroad} and \ref{dilptmultbroad}, we show our 
predicted dilepton invariant mass and transverse momentum spectrum for 
RHIC and LHC energies assuming the time dependence of the energy 
density, the hard momentum scale and the anisotropy parameter are 
given by Eqns.~(\ref{eq:modelEQs}) and (\ref{UdeffFixMult}) with 
$\delta=2/3$. This corresponds to our collisionally-broadened 
interpolating model with fixed final multiplicity.  In 
Fig.~\ref{enhance1-mult} we show the dilepton enhancement, $\phi$, as 
function of transverse momentum for $\tau_{\rm iso} = 2$ fm/c at 
(left) RHIC energies (right) LHC energies.  The invariant mass cut is 
the same as in Fig.~\ref{dilptmultbroad} ($0.5\,\leq\,M\,\leq\,1$ 
GeV).  As can be seen from Fig.~\ref{enhance1-mult} similar to the 
case of fixed initial conditions there is a rapid increase in $\phi$ 
between 1 and 3 GeV at RHIC energies and 1 and 4 GeV at LHC energies. 
However, compared to the case of the collisionally-broadened 
interpolating model with fixed initial condition (Fig.~\ref{enhance}) 
the maximum enhancement is reduced slightly and we see a more 
pronounced peak in $\phi$ as a function of transverse momentum 
appearing. As can be seen from this figure both sharp and smooth 
transitions from early-time collisionally-broadened expansion to ideal 
hydrodynamic expansion result in a 20-40\% enhancement of medium 
dilepton yields at RHIC energies, and 30-50\% at LHC energies.

\subsection{Summary of Results}
\label {sec:resultssummary}

Based on the figures presented in the previous subsections we see that 
the best opportunity for measuring information about plasma initial 
conditions is from the $M < 2$ GeV dilepton transverse momentum 
spectra between 1 $< p_T < $ 6 GeV at RHIC and 2 $< p_T 
<$ 8 GeV at LHC. This is due to the fact that medium dilepton 
yields dominate other mechanisms in that kinematic range and hence 
give the cleanest possible information about plasma initial 
conditions.  In all cases shown above dilepton production is enhanced 
by pre-equilibrium emissions with the largest enhancements occurring 
when assuming fixed initial conditions and the free-streaming 
interpolating model.  As we have mentioned above this model sets the 
upper-bound for the expected dilepton enhancement.  Our most 
physically realistic model is the collisionally-broadened 
interpolating model with fixed final multiplicity so we will use it 
for our final predictions of expected dilepton enhancement.  For this 
model, as can be seen from Fig.~\ref{enhance1-mult}, assuming 
$\tau_{\rm iso} = 2$ fm/c we find a 20-40\% enhancement in dilepton 
yields at RHIC and 30-50\% at LHC.

\begin{figure*}[t]
\includegraphics[width=8.1cm]{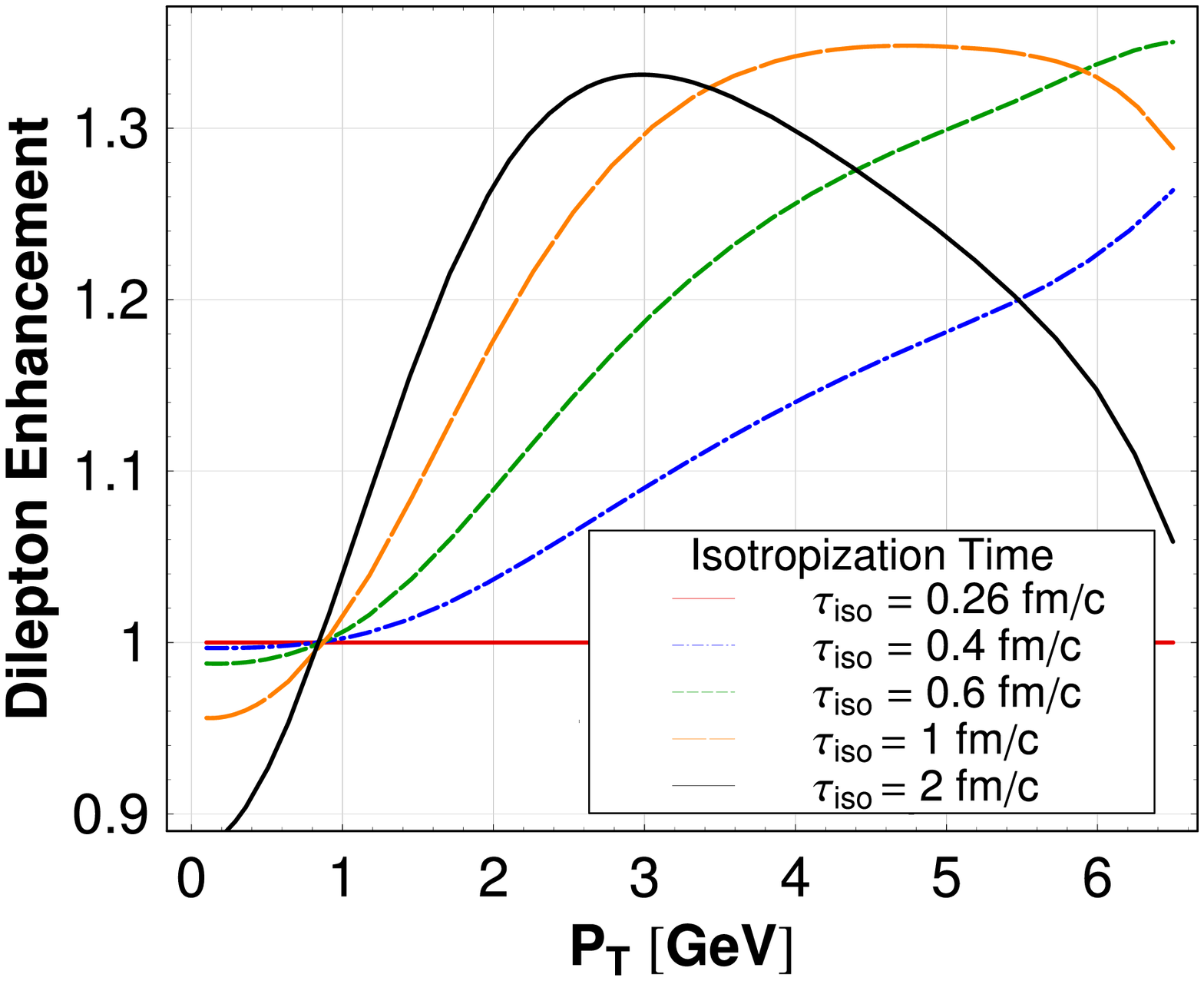}\quad\quad
\includegraphics[width=8.8cm]{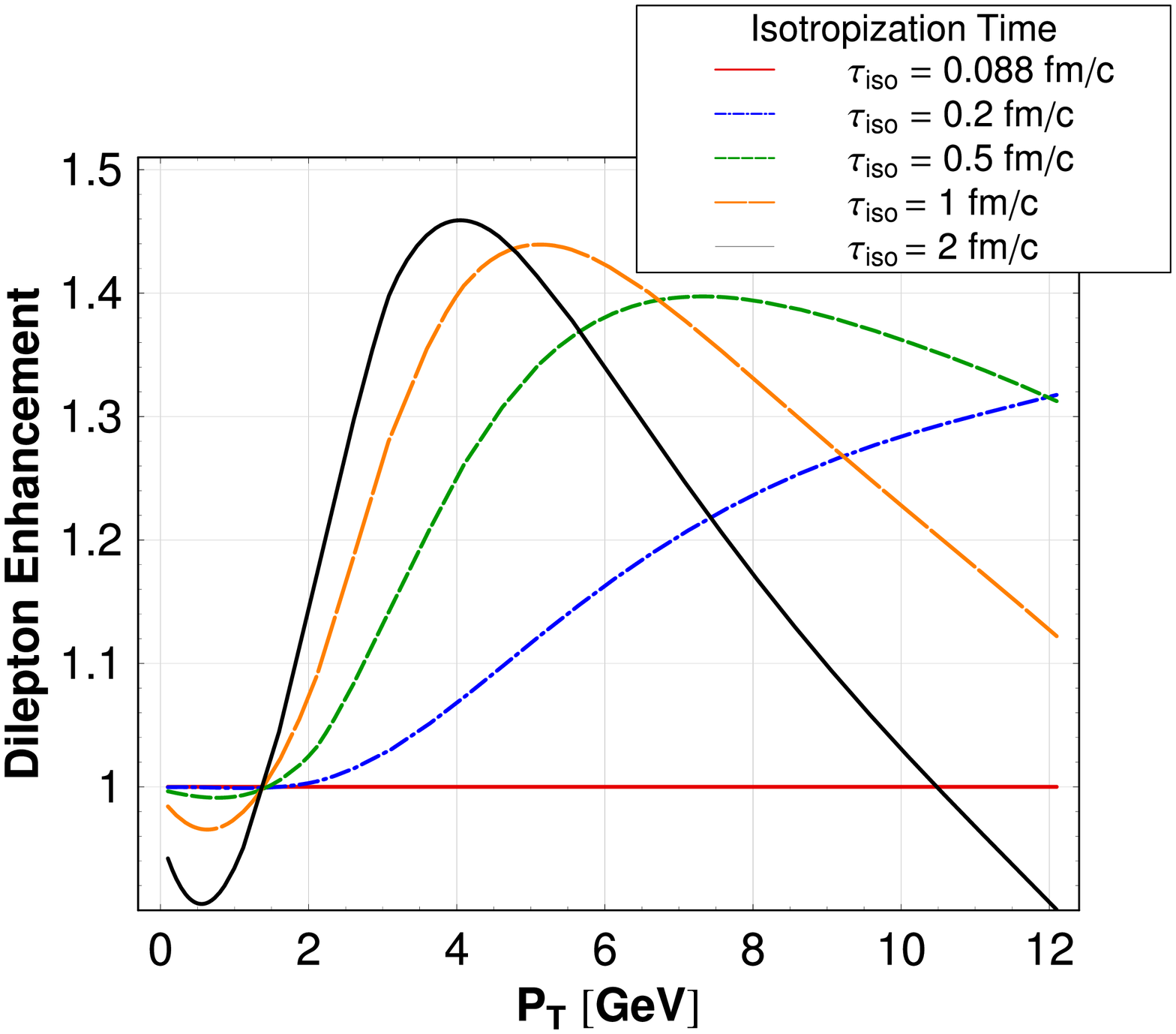}
\vspace{-2mm}
\caption{
Dilepton enhancement, $\phi$, as defined in Eq.~(\ref{dileptonenhancement}) 
resulting from our collisionally-broadened interpolating model 
($\delta=2/3$) with fixed final multiplicity and $\gamma=2$.  
Result for RHIC energies is shown on left and for LHC energies 
on right.  The invariant mass cut used was $0.5\,\leq\,M\,\leq\,1$ GeV and and rapidity $y$=0.
Lines show expected pre-equilibrium dilepton enhancements 
for different values of the assumed plasma isotropization time, $\tau_{\rm iso}$.
}
\label{enhance2-mult}
\end{figure*}

In addition we can calculate the dilepton enhancement for different 
assumed values for $\tau_{\rm iso}$.  This is shown for RHIC energies 
(left) and LHC energies (right) in Fig.~\ref{enhance2-mult} where we 
have fixed $\gamma=2$ and varied $\tau_{\rm iso}$ to see the effect of 
varying the assumed isotropization time.  As can be seen from this 
figure the effect of reducing $\tau_{\rm iso}$ is to shift the peak in 
$\phi$ to larger $p_T$ while at the same time reducing the overall 
amplitude of the peak.  This feature seems generic at both RHIC and 
LHC energies.  Therefore, in order to see the difference between an 
instantaneously thermalized QGP with $\tau_{\rm iso} = \tau_0$ and one 
with a later thermalization time requires determining the medium 
dilepton spectra between $1 < p_T < 6$ GeV at RHIC and $2 < p_T < 8$ 
GeV at LHC with high precision so that one could measure the less than 
50\% variation resulting from pre-equilibrium emissions.

\begin{figure*}[t]
\includegraphics[width=8.4cm]{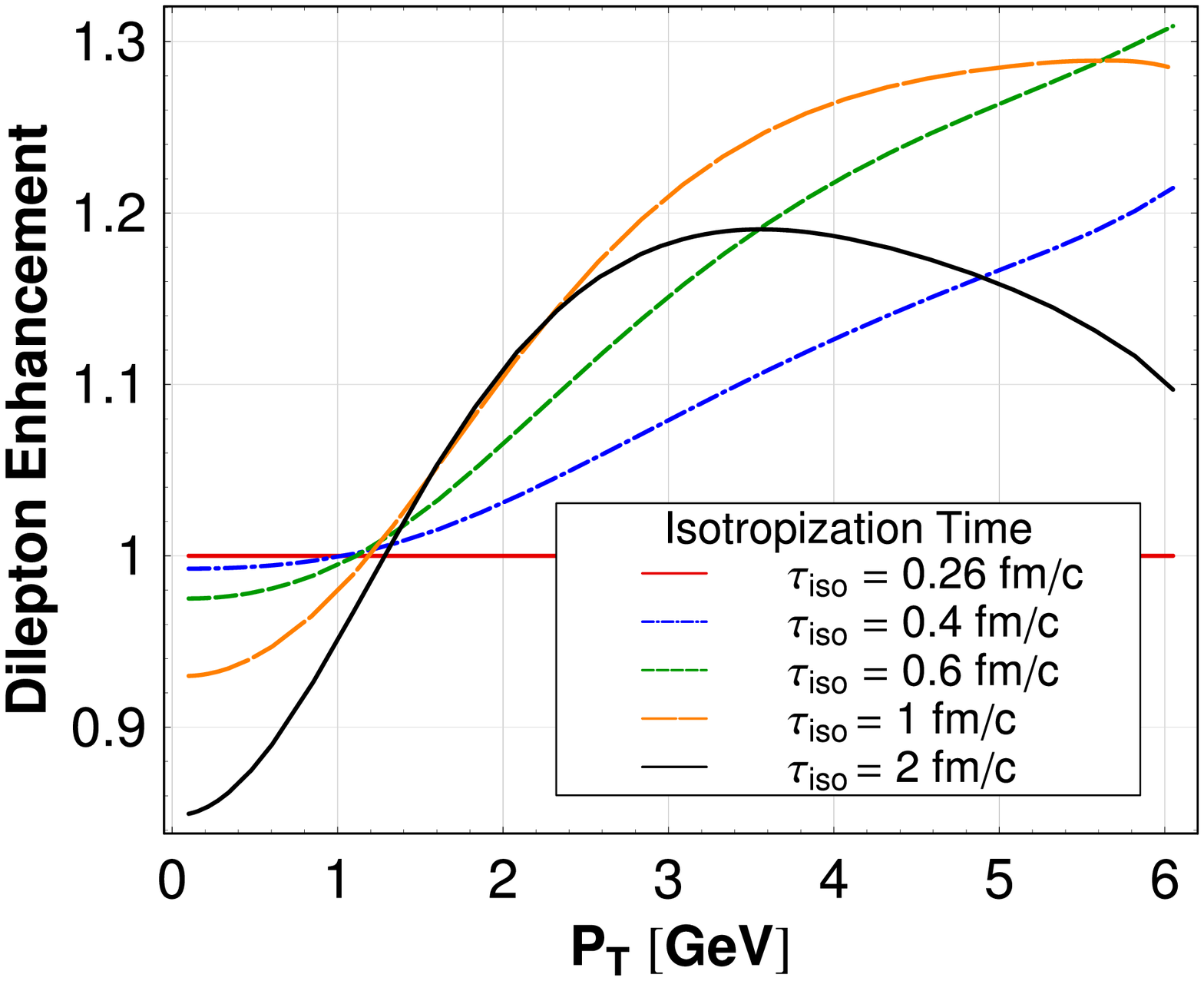}\quad\quad
\includegraphics[width=8.4cm]{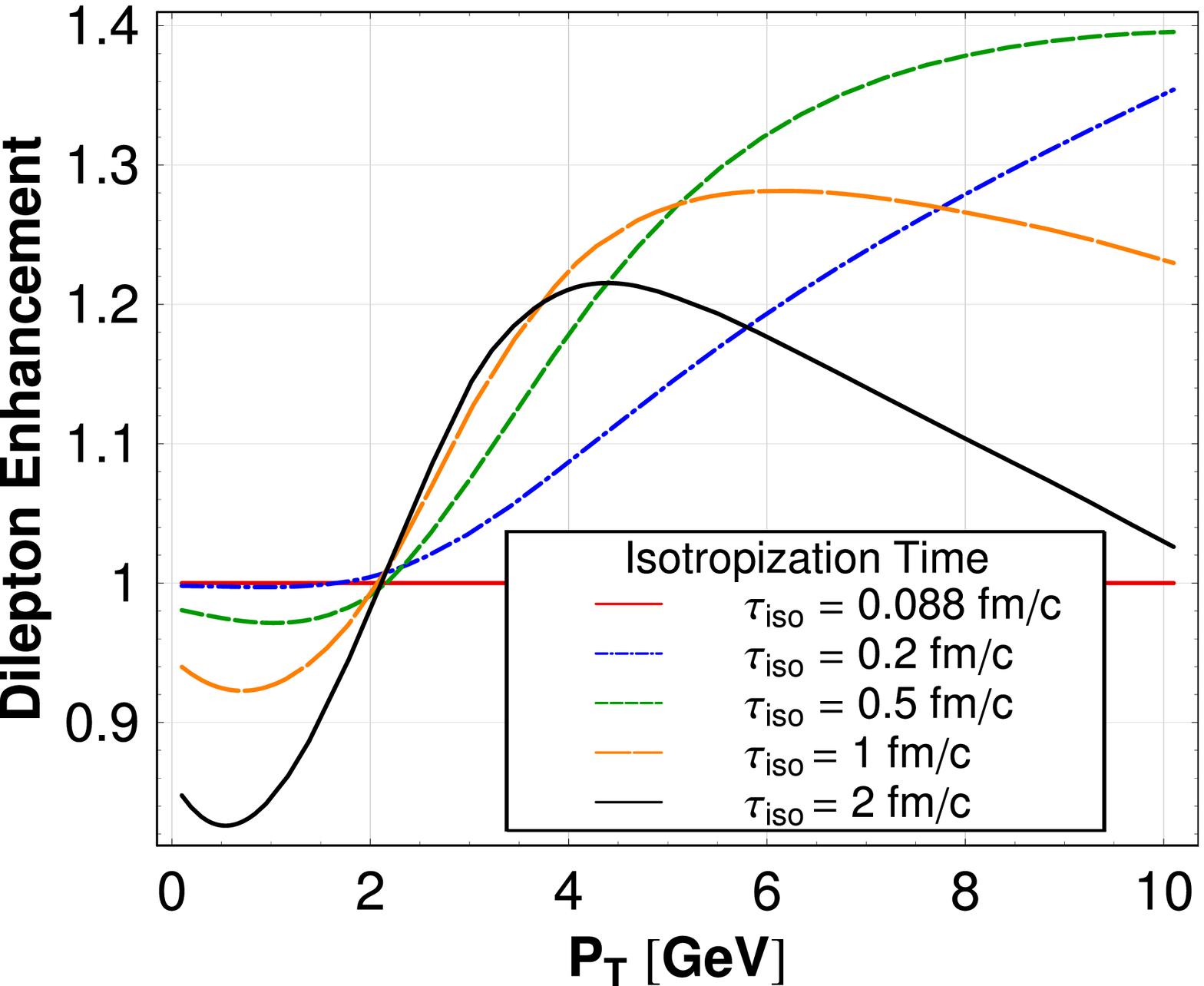}
\vspace{-2mm}
\caption{
Dilepton enhancement, $\phi$, as defined in Eq.~(\ref{dileptonenhancement}) 
resulting from our collisionally-broadened interpolating model 
($\delta=2/3$) with fixed final multiplicity and $\gamma=2$.  
Result for RHIC energies is shown on left and for LHC energies 
on right.  The invariant mass cut used was $1\,\leq\,M\,\leq\,2$ GeV and and rapidity $y$=0.
Lines show expected pre-equilibrium dilepton enhancements 
for different values of the assumed plasma isotropization time, $\tau_{\rm iso}$.
}
\label{enhance2-mult-cut2}
\end{figure*}

Finally, we point out that in Fig.~\ref{enhance2-mult} we have chosen an 
invariant mass cut of $0.5 < M < 1$ GeV.  Since our model predicts the 
full yields versus $M$ and $p_T$ it is possible to take other cuts 
(invariant mass and/or transverse momentum).  This could be coupled 
with fits to experimental data, allowing one to fix $\tau_{\rm iso}$ 
and $\gamma$ via a ``multiresolution'' analysis.  To demonstrate the 
dependence of $\phi$ on the mass cut in Fig.~\ref{enhance2-mult-cut2} 
we show the dilepton enhancement, $\phi$, using a mass cut of $1 < M < 
2$ GeV. As can be seen from this Figure the qualitative features of 
our model's predictions are similar to the lower mass cut presented in 
Fig.~\ref{enhance2-mult}; however, for this mass cut we see that there 
is a stronger suppression of dilepton production at low and high 
invariant masses if there is late thermalization, $\tau_{\rm iso} 
\gsim 2$ fm/c.  Such features can be used to constrain the model 
further when confronted with experimental data.

\section{Conclusions}
\label {sec:conclusion}

In this work we have presented models which allow one to smoothly 
interpolate between early-time non-equilibrium 1+1 dimensional 
expansion to late-time isotropic equilibrium 1+1 dimensional 
hydrodynamic expansion. To accomplish this we introduced simple 
interpolating models with two parameters: $\tau_{\rm iso}$, which is 
the time at which the system begins to expand hydrodynamically and 
$\gamma$ which sets the width of the transition.  Using these models 
we integrated the leading order rate for dilepton production in an 
anisotropic plasma over our modeled space-time evolution.  Based on 
our numerical results for the variation of dilepton yields with the 
assumed values of $\tau_{\rm iso}$ we find that the best opportunity 
to determine information about the plasma isotropization time is by 
analyzing the high transverse momentum (1 $< p_T < $ 6 GeV at RHIC and 
2 $< p_T <$ 8 GeV at LHC) dilepton spectra using relatively low pair 
invariant mass cuts ($M \lsim 2$ GeV).  Based on these $p_T$ spectra 
we introduced the ``dilepton enhancement'' factor $\phi(\tau_{\rm 
iso})$ which measures the ratio of yields obtained from a plasma which 
isotropizes at $\tau_{\rm iso}$ to one which isotropizes at the 
formation time, $\tau_0$.  

We showed that for our most extreme model, the free-streaming 
interpolating model ($\delta=2$) with fixed initial conditions, that 
the resulting enhancement $\phi$ can be as large as 10; however, this 
extreme model probably overestimates the amount of anisotropy in the 
plasma.  Additionally, this model results in a large amount of entropy 
generation during the transition from the free-streaming $\tau^{-1}$ 
asymptotic behavior to hydro $\tau^{-4/3}$ asymptotic behavior.  As we 
discussed this greatly constrains the maximum isotropization times 
$\tau_{\rm iso}$ which are consistent with experimental indications of 
low (10-20\%) entropy generation.

In order to construct a more realistic model we then included 
collisional-broadening of the initial pre-equilibrium parton 
distribution functions ($\delta=2/3$).  In this more realistic model 
there is much less entropy generation and the system is always closer 
to ideal 1+1 hydrodynamic expansion than in the free-streaming 
interpolating model.  As a result the dilepton enhancement due to 
pre-equilibrium emissions is lower than the free-streaming case.  We 
find that when fixing final multiplicity at RHIC energies there is a 
20-40\% enhancement in the high-transverse momentum dileptons and at 
LHC energies it is 30-50\% when one assumes an isotropization time of 
$\tau_{\rm iso} = 2$ fm/c.  The amplitude of the enhancement and 
position of the peak in the enhancement function, $\phi$, varies with 
the assumed value of $\tau_{\rm iso}$ which, given sufficiently 
precise data, would provide a way to determine the plasma 
isotropization time experimentally.  We presented our predictions for 
the dilepton enhancement, $\phi$, as a function of $\tau_{\rm iso}$ 
for two different invariant mass cuts, demonstrating that our model 
can be constrained by a multiresolution analysis which should 
give higher statistics and further constrain the two model parameters 
at our disposal.

One shortcoming of this work is that we haven't included NLO 
corrections to dilepton production from an anisotropic QGP.  At low 
invariant mass these corrections would become important.  As a next 
step one must undertake a calculation of the rate for dilepton pair 
production at NLO in an anisotropic plasma.  This is complicated by the 
presence of plasma instabilities which render some expressions like 
$\langle AA \rangle$ correlators formally divergent and hence 
analytically meaningless.  However, when combined with numerical 
solution of the long-time behavior of a plasma subject to the 
chromo-Weibel instability it may be possible to extract finite 
correlators \cite{ArnoldPersonal}.  This is a daunting but doable task. 
Absent such a calculation, phenomenologically speaking it is probably a 
very good approximation to simply take existing NLO calculations and 
apply the enhancement function $\phi$ as calculated at LO.  We leave 
this for future work.

Another uncertainty comes from our implicit assumption of chemical 
equilibrium. If the system is not in chemical equilibrium (too many 
gluons and/or too few quarks) early time quark chemical potentials, 
or fugacities, will affect the production of lepton pairs 
\cite{Dumitru:1993vz,Strickland:1994rf}.  However, to leading order 
the quark and gluon fugacities will cancel between numerator and 
denominator in the dilepton enhancement, $\phi$ 
\cite{Strickland:1994rf}.  We, therefore, expect that to good 
approximation one can factorize the effects of momentum space 
anisotropies and chemical non-equilibrium.

We note in closing that the interpolating model presented here has 
application beyond the realm of computing dilepton yields.  In fact, 
such a model can be used to assess the phenomenological consequences 
of momentum-space anisotropies in other possible observables which are 
sensitive to early-time stages of the QGP, e.g., photon production \cite{Schenke:2006yp}, 
heavy-quark transport, jet-medium induced electromagnetic radiation, 
etc.


\begin{acknowledgments}

We thank to A. Dumitru, M. Gyulassy, A. Ipp, A. Rebhan, and B. Schenke 
for helpful discussions. We also thank S. Turbide for providing us 
predictions for the other relevant sources of dilepton production. M. 
Martinez gratefully acknowledges support by the Helmholtz Research 
School and Otto Stern School of the Johann Wolfgang 
Goethe-Universit\"at. M.S. was supported by DFG project GR 1536/6-1.

\end{acknowledgments}


\appendix






\begin {thebibliography}{}

\bibitem{Huovinen:2001cy}
  P.~Huovinen, P.~F.~Kolb, U.~W.~Heinz, P.~V.~Ruuskanen and S.~A.~Voloshin,
  Phys.\ Lett.\  B {\bf 503} (2001) 58.

\bibitem{Hirano:2002ds}
  T.~Hirano and K.~Tsuda,
  Phys.\ Rev.\  C {\bf 66} (2002) 054905.

\bibitem{Tannenbaum:2006ch}
  M.~J.~Tannenbaum,
  Rept.\ Prog.\ Phys.\  {\bf 69} (2006) 2005.

\bibitem{Luzum:2008cw}
  M.~Luzum and P.~Romatschke,
  Phys.\ Rev.\  C {\bf 78}, 034915 (2008).

\bibitem{Jas:2007rw}
  W.~Jas and S.~Mrowczynski,
  Phys.\ Rev.\  C {\bf 76}, 044905 (2007).
  
\bibitem{Baier:2000sb}
  R.~Baier, A.~H.~Mueller, D.~Schiff and D.~T.~Son,
  Phys.\ Lett.\  B {\bf 502} (2001) 51.

\bibitem{Strickland:2007fm}
  M.~Strickland,
  J.\ Phys.\ G {\bf 34}, S429 (2007).
  
\bibitem{Mrowczynski:2000ed}
  S.~Mrowczynski and M.~H.~Thoma,
  Phys.\ Rev.\  D {\bf 62} (2000) 036011.

\bibitem{Randrup:2003cw}
  J.~Randrup and S.~Mrowczynski,
  Phys.\ Rev.\  C {\bf 68} (2003) 034909.

\bibitem{Romatschke:2003ms}
  P.~Romatschke and M.~Strickland,
  Phys.\ Rev.\  D {\bf 68} (2003) 036004.

\bibitem{Romatschke:2004jh}
  P.~Romatschke and M.~Strickland,
  Phys.\ Rev.\  D {\bf 70}, 116006 (2004).
  
\bibitem{Arnold:2003rq}
  P.~Arnold, J.~Lenaghan and G.~D.~Moore,
  JHEP {\bf 0308} (2003) 002.

\bibitem{Mrowczynski:2004kv}
  S.~Mrowczynski, A.~Rebhan and M.~Strickland,
  Phys.\ Rev.\  D {\bf 70} (2004) 025004.

\bibitem{Arnold:2004ti}
  P.~Arnold, J.~Lenaghan, G.~D.~Moore and L.~G.~Yaffe,
  Phys.\ Rev.\ Lett.\  {\bf 94} (2005) 072302.
  
\bibitem{Rebhan:2004ur}
  A.~Rebhan, P.~Romatschke and M.~Strickland,
  Phys.\ Rev.\ Lett.\  {\bf 94} (2005) 102303.

\bibitem{Arnold:2005vb}
  P.~Arnold, G.~D.~Moore and L.~G.~Yaffe,
  Phys.\ Rev.\  D {\bf 72} (2005) 054003.

\bibitem{Rebhan:2005re}
  A.~Rebhan, P.~Romatschke and M.~Strickland,
  JHEP {\bf 0509} (2005) 041.

\bibitem{Romatschke:2005pm}
  P.~Romatschke and R.~Venugopalan,
  Phys.\ Rev.\ Lett.\  {\bf 96} (2006) 062302.

\bibitem{Schenke:2006xu}
  B.~Schenke, M.~Strickland, C.~Greiner and M.~H.~Thoma,
  Phys.\ Rev.\  D {\bf 73} (2006) 125004.

\bibitem{Schenke:2006fz}
  B.~Schenke and M.~Strickland,
  Phys.\ Rev.\  D {\bf 74} (2006) 065004.

\bibitem{Manuel:2006hg}
  C.~Manuel and S.~Mrowczynski,
  Phys.\ Rev.\  D {\bf 74} (2006) 105003.

\bibitem{Romatschke:2006nk}
  P.~Romatschke and R.~Venugopalan,
  Phys.\ Rev.\  D {\bf 74} (2006) 045011.

\bibitem{Bodeker:2007fw}
  D.~Bodeker and K.~Rummukainen,
  JHEP {\bf 0707} (2007) 022.
  
\bibitem{Romatschke:2006wg}
  P.~Romatschke and A.~Rebhan,
  Phys.\ Rev.\ Lett.\  {\bf 97} (2006) 252301.

\bibitem{Dumitru:2005gp}
  A.~Dumitru and Y.~Nara,
  Phys.\ Lett.\  B {\bf 621} (2005) 89.

\bibitem{Dumitru:2006pz}
  A.~Dumitru, Y.~Nara and M.~Strickland,
  Phys.\ Rev.\  D {\bf 75} (2007) 025016.

\bibitem{Rebhan:2008uj}
  A.~Rebhan, M.~Strickland and M.~Attems,
  Phys.\ Rev.\  D {\bf 78}, 045023 (2008).

\bibitem{Gale:2003iz}
  C.~Gale and K.~L.~Haglin,
  arXiv:hep-ph/0306098.

\bibitem{Turbide:2006zz}
  S.~Turbide,
  ``Electromagnetic radiation from matter under extreme conditions'', PhD Dissertation, UMI-NR-25272 (2006).
  

\bibitem{Mauricio:2007vz}
  M.~Martinez and M.~Strickland,
  Phys.\ Rev.\ Lett.\  {\bf 100} (2008) 102301.

\bibitem{Kapusta:1992uy}
  J.~I.~Kapusta, L.~D.~McLerran and D.~Kumar Srivastava,
  Phys.\ Lett.\  B {\bf 283}, 145 (1992).

\bibitem{Dumitru:1993vz}
 A.~Dumitru, D.~H.~Rischke, T.~Schonfeld, L.~Winckelmann, H.~Stocker and W.~Greiner,
  Phys.\ Rev.\ Lett.\  {\bf 70} (1993) 2860.
  
\bibitem{Strickland:1994rf}
  M.~Strickland,
  Phys.\ Lett.\  B {\bf 331}, 245 (1994).

\bibitem{Ollitrault:2007du}
  J.~Y.~Ollitrault,
  Eur.\ J.\ Phys.\  {\bf 29} (2008) 275.

\bibitem{Thoma:1997dk}
  M.~H.~Thoma and C.~T.~Traxler,
  Phys.\ Rev.\  D {\bf 56} (1997) 198.

\bibitem{Arnold:2002ja}
  P.~Arnold, G.~D.~Moore and L.~G.~Yaffe,
  JHEP {\bf 0206} (2002) 030.

\bibitem{Arleo:2004gn}
  F.~Arleo {\it et al.},
  arXiv:hep-ph/0311131.

\bibitem{Turbide:2006mc}
  S.~Turbide, C.~Gale, D.~K.~Srivastava and R.~J.~Fries,
  Phys.\ Rev.\  C {\bf 74} (2006) 014903.

\bibitem{Kajantie:1986dh}
  K.~Kajantie, J.~I.~Kapusta, L.~D.~McLerran and A.~Mekjian,
  Phys.\ Rev.\  D {\bf 34}, 2746 (1986).

\bibitem{Kampfer:1992bb}
  B.~Kampfer and O.~P.~Pavlenko,
  Phys.\ Lett.\  B {\bf 289}, 127 (1992).

\bibitem{Bjorken:1982qr}
  J.~D.~Bjorken,
  Phys.\ Rev.\  D {\bf 27}, 140 (1983).
  
\bibitem{Baym:1984np}
  G.~Baym,
  Phys.\ Lett.\  B {\bf 138}, 18 (1984).

\bibitem{Dusling:2008xj}
  K.~Dusling and S.~Lin,
  Nucl.\ Phys.\  A {\bf 809} (2008) 246.

\bibitem{Bodeker:2005nv}
  D.~Bodeker,
  JHEP {\bf 0510} (2005) 092.
  
\bibitem{Arnold:2005qs}
  P.~Arnold and G.~D.~Moore,
  Phys.\ Rev.\  D {\bf 73} (2006) 025013.
  
\bibitem{Arnold:2007cg}
  P.~Arnold and G.~D.~Moore,
  Phys.\ Rev.\  D {\bf 76} (2007) 045009.

\bibitem{Dumitru:2007qr}
  A.~Dumitru, E.~Molnar and Y.~Nara,
  Phys.\ Rev.\  C {\bf 76}, 024910 (2007).

\bibitem{AndiPersonal}
	A.~Ipp, personal communication.

\bibitem{ArnoldPersonal}
	P.~Arnold, personal communication.

\bibitem{Schenke:2006yp}
  B.~Schenke and M.~Strickland,
  Phys.\ Rev.\  D {\bf 76}, 025023 (2007).
	      
\end {thebibliography}
\end {document}